\documentclass[12pt]{article} 
\usepackage{authblk}
\usepackage{array}
\usepackage{amsmath}
\usepackage{amssymb}
\usepackage[english]{babel}
\usepackage[letterpaper, top=2cm, bottom=2cm, left=3cm, right=3cm, marginparwidth=1.75cm]{geometry}
\usepackage{amsmath,amssymb,booktabs,tabularx,algorithm,algorithmicx,algpseudocode} 
\usepackage[colorlinks=true, allcolors=blue]{hyperref}
\usepackage{natbib}
\usepackage{tikz}
\usetikzlibrary{arrows.meta,positioning}

\usepackage{tikz}
\usetikzlibrary{shapes, decorations.markings, graphs, quotes, arrows.meta}
\usepackage[european,s traightvoltages, RPvoltages, americanresistor, americaninductors]{circuitikz}
\ctikzset{bipoles/thickness=1.2, label distance=1mm} 

\usepackage{amsmath}

\usepackage{caption}
\usepackage{subcaption}
\usepackage{graphicx}

\usepackage{hyphenat}
\usepackage[loadshadowlibrary,textsize=tiny]{todonotes}
\todostyle{tobia}{color=teal!50, shadow}
\todostyle{emilio}{color=red!50, shadow}
\todostyle{claudio}{color=brown!50, shadow}

\renewcommand{\arraystretch}{1.35}

\title{A Triad of Networks and a Triad of Fusions for the {\em Other Climate Crisis}}
\author[1,3]{Emilio Porcu}
\author[2]{Tobia Filosi}
\author[3]{Horst Simon}
\affil[1]{Khalifa University, Department of Mathematics}
\affil[2]{University of Trento, Department of Mathematics}
\affil[3]{ADIA Lab}

\begin{document}
\maketitle

\tableofcontents \newpage 

\begin{abstract}
    \citeauthor{ShawStevens2025} call for a new paradigm in climate science criticizes Large Scale Determinism in favor of (i) embracing discrepancies, (ii) embracing hierarchies, and (iii) create disruption while keeping interpretability. \par
    The last 20 years have seen a plethora of contributions relating complex networks with climate data and climate models. We provide a view of climate networks through a triad of frameworks and associated paradigms: (a) networks {\em of} data, where both (geographical) nodes and their links (arcs) are determined according to some metrics and/or statistical criteria; (b) climate data {\em over} networks, where the structure of the network (for both vertices and edges) is topologically pre-determined, and the climate variable is continuously defined over the (nonlinear) network; finally, (c) networks {\em for} data, referring to the huge machinery based on networks within the realm machine learning and statistics, with specific emphasis on their use for climate data. \par    This paper is not a mere description of each element of the network triad, but rather a manifesto for the creation of three classes of fusions (we term them {\em bridges}). We advocate and carefully justify a fusion {\em within} to provide a {\em corpus unicuum} inside the network triad. We then prove that the fusion {\em within} is the starting point for a fusion {\em between}, where the network triad becomes a condition {\em sine qua non} for the implementation of the Shaw-Stevens agenda. We culminate with a {\em meta} fusion that allows for the creation of what we term a Shaw-Stevens network ecosystem. 
    
    \bigskip    
    
\end{abstract}

\newpage

\section*{Preludium: the {\em Other Climate Crisis}}
\addcontentsline{toc}{section}{Preludium: the {\em Other Climate Crisis}}

The recent \emph{tour de force} by \cite{ShawStevens2025} has caused an intense debate about what climate science can predict and how those predictions are to be justified. While \citeauthor{ShawStevens2023} do not question the law of physics to explain global warming, they warn that the paradigm of Large Scale Determinism (LSD) -- being the dominant culture in climate prediction -- is under strain when analyzed within the realm of regional phenomena having direct societal relevance. \par
According to the LSD paradigm, small–scale processes (deep convection, boundary layer turbulence, ocean eddies) are parameterized as functions of large–scale state variables. The LSD has been a major breakthrough in climate science, and has allowed for a better understanding of certain dynamics, from the attribution of anthropogenic warming to a broad set of large–scale signals. However, it is being questioned because of the increasing amount of discrepancies at regional scales where local phenomena pop up. Rephrasing, \citeauthor{ShawStevens2023} note that, as observations lengthen and as regional signals sharpen, the LSD hypothesis encounters accumulating anomalies, which translate into large discrepancies between what is simulated on the basis of LSD and what is locally attained. The most challenging regimes are those where scale coupling is strongest, such as, for instance, the deep tropics, the Southern Ocean, and high–impact extremes. Additionally, there might be uncertainties in boundary conditions and structural choices that observations cannot yet tightly constrain. In conclusion, \citeauthor{ShawStevens2023} claim that this is the “Other Climate Crisis” of methodology and predictability.  \par
\citeauthor{ShawStevens2023} do not deny the established agenda. However, they constructively criticize the existing paradigm and integrate it with (a) {embracing discrepancies as tests}, (b) {resort to hierarchies}, and (c) {make experiments with disruptive tools} \citep{ShawStevens2025}. As for (a), embracing discrepancies allows to treat mismatches as opportunities to falsify 
 mechanistic claims, and to design observations that actually adjudicate between competing hypotheses. Regarding (b), hierarchies are a natural instrument to provide fusions in the sense of components
 (atmosphere, ocean, land) and scales (from convection to planetary waves), so that complexity does not obfuscate interpretability. Finally, disruption (c) can be achieved in three directions: very large ensembles for an improved uncertainty accuracy; global models run at kilometer resolution; and machine learning, both as a tool as well as a new paradigm for building predictive emulators. According to the authors, weather predictability has a mature theory, and climate predictability does not. Understanding how the future looks like under these new paradigms is a major scientific task for the coming decade \citep{ShawStevens2025}. \par

The perspective provided in the present paper (a triad of networks for climate data and a triad of fusions) stems from Shaw-Stevens motivation while deviating ontologically, and while taking a complementary route. Our point is that our organization of networks and consequent paradigm provides a concrete support for the implementation of Shaw-Stevens agenda. The way this will happen is described throughout, and we avoid anticipating too much in this direction.  And there is more: we show how what we term fusion {\em within} the proposed network triad, and then {\em between} the triad and the Shaw-Stevens agenda, becomes then the natural environment of what we term a {\em meta} fusion that opens the door to the upcoming Shaw-Stevens networks ecosystem. \par 
The remainder of the paper develops these claims.

\section{Introduction}

\subsection{Context}

The word \emph{network} is everywhere in the entire landscape of contemporary data science and artificial intelligence.  
At the methodological level, the word {\em network} is reminiscent to the architectures that are the basis of the transformation of modern learning -- starting with the densely connected layers of neural networks, until the graph based operators of geometric deep learning.  
At the applied level, {\em network} is a natural way of thinking when the goal is to describe the structures through which complex systems are represented: social networks for interaction, transportation networks for flow, and climate networks for interdependence, to mention a few. \par 
Hence, the word {\em networks} transcended their canonical meaning to become a universal language for expressing relation, propagation, and emergence. So much so, that we find difficult to talk about data science or AI without speaking about {\em networks}. \par 
Climate (data) science is certainly an area where networks are ubiquitous. Network-based methods integrate both data-driven and model-based tools to uncover underlying structures in climate dynamics, particularly in complex space-time systems. This paper  introduces a triad of networks that serves as a unifying framework for interpreting climate data. The paper is not a descriptive and aesthetics exercise, as will be shown throughout the subsequent sections. 
{\renewcommand{\labelenumi}{(\alph{enumi})}
\begin{enumerate}
    \item Networks {\em of} Climate Data: here, we start from climate data (typically, climate model outputs) and construct a network where the nodes are the spatial locations, and where the edges are a graphical representation of certain statistical dependencies. For example: {\em if the temporal correlation between a time series at location $\boldsymbol{x}_1$ and another at location $\boldsymbol{x}_2$ is bigger then $0.5$, then we draw an edge connecting them.}
    \item Climate Data {\em over} Networks: in this case, the network structure is not inferred through some statistical procedures, but given. It is a representation of a spatial, or a physical, or a topological structure ({\em e.g.}, a river basin, transport grid), and climate data (e.g., temperature, precipitation) are modeled or predicted as processes that are continuously indexed over the nodes and edges of this network.
    \item Climate Data {\em for Networks}: modeling approaches based on networks, that are typically coming from machine learning and AI, and are used to understand the structure of climate data (unsupervised learning) or predict their patterns (supervised).
\end{enumerate}}
These classes are not mutually exclusive: there are obvious intersections, especially the pairs (a,c) and (b,c). Less obvious is the intersection (a,b) -- however, this inspires considerable amount of research, new methods, and climate applications for the future. \par
The juxtaposition of these paradigms serves as a platform for providing unifying perspectives, highlight methodological advances, challenges, and provide opportunities to integrate different paradigms and philosophies in climate data science. \par
Table \ref{tab:comparativeDescription} opens for some preliminary considerations about the difference between the first two paradigms.
\begin{table}[ht] 
    \centering
    \resizebox{\textwidth}{!}{
        \begin{tabular}{|c|c|c|}
        \hline
            {\bf Feature} & {\bf Networks of Climate Data} & {\bf Climate Data over Networks}  \\
            \hline 
            Network Source  &  Inferred from Data & Pre-Determined \\
            Node  & Spatial Location & Locations in a Structured Network \\
            Edge means  & Statistically Similar ({Synchronization})  & Physical connection \\
            Covariance & Through similarity  & Direct constructions \\
            Typical Applications  & Telecommunication Analysis, Flipping Points & Kriging \\
            \hline
        \end{tabular}
    }
    \caption{A comparative description of networks {\em of} and {\em for} through some features.}
    \label{tab:comparativeDescription}
\end{table}
Clearly, the first two paradigms are coming from similar frameworks, as the reference set is, for both, space-time. Hence, the space-time correlation plays an important role for both. Robustness from both perspective of noisy data and models covers a central aspect for both paradigms. A recent contribution \citep{haas2023pitfalls} proves how important can be sampling errors for the first paradigm. \par
The three paradigms are compatible, and their fusion and bridging will cover an important part of the subsequent sections. \par
There are clear intersections (pairwise and triplewise): {\em of} and {\em for} have an obvious intersection as the networks extracted from data are then used as analysis tools -- for instance, feedback into models, or guiding assimilation strategies. As for {\em of} and {\em over}, the topologies extracted from data can be then used as ``spatial" domains where some processes are continuously defined. For instance, propagation of certain phenomena in climate dynamics. {\em For} and {\em over} have an obvious intersection - for instance, the topology created through the first might be embedded in the second to build an hybrid structure. Finally, the three of them intersect through the sequence {\em of}-{\em over}-{\em for}. For instance, dynamic climate networks used in spatiotemporal modeling over changing graphs. \par

\subsection{Networks {\em of}: Some Literature}
The interpretation of the climate system as a network emerged after the painstaking work by Prof. A. Tsonis and his group. The work by \citet{tsonis2006networks} provides a paradigm where geographic locations or climate indices are nodes, where their statistical affinities are edges. \par
Tsonis' paradigm provided a simple construction that fixed a common language for teleconnections, allowing correlations, mutual information, or event synchronization to be re-expressed as a topological structure. \par
Subsequent developments and findings extended the paradigm through the inclusion of questions related with architecture and organization \citep[see the characterization of climate network topology in][]{tsonis2008topology, Tsonis2004}. \par
While Tsonis paradigm has been the building block, subsequent contributions have provided more sophistication inside a very promising architecture. This new wave has gone under the trendy word of {\em complex networks theory}: 
\cite{Donges2009EPL} introduced the notion of a basis network to reduce redundancy and emphasize the most informative connections, while \cite{donges2009complex} provided a broader methodological framework built upon linking climate dependence measures with canonical network diagnostics (degree, clustering, centrality, modularity). \par
Introducing temporal evolution within complex systems is a natural step. Evolving networks allow, for instance, to better understand different types of El Ni{\~n}o episodes as they allow to track changes in connectivity patterns \citep{radebach2013disentangling}.  By quantifying structural changes in connectivity, network measures allow an objective separation of Eastern- and Central-Pacific El Niño/La Niña episodes \citep{wiedermann2016climate}. \par
Networks {\em for} have been proved useful to show that connectivity among rare events (extreme values) can reveal long-range structures that would result undetected with classical methodologies (for instance, Gaussian processes) as shown for Indian Summer Monsoon precipitation \citep{stolbova2014topology} and for global extreme-rainfall teleconnections \citep{boers2019complex}. \par 
Stability, embedding, and prediction have also played a major role regarding sophistication of the basic networks structure. 
\citet{berezin2012stability} studied persistence and stability in climate networks. They proved that much of their apparent robustness is due to the geographical embedding of nodes in concert with the physical coupling among different regions of the climate system. 
Predictive analysis has drown the attention of several scientists: \cite{ludescher2013improved} reported improvements in El Niño forecasting via cooperativity detection, while interaction-network early-warning indicators were proposed for a potential AMOC collapse \citep{van2013interaction} and further consolidated through observation-based signals of AMOC weakening \citep{Boers2021}. \par
All the previously-mentioned contributions are mostly based on {\em ad hoc} procedures rather than statistical rigor. This has come into place more recently, with \cite{deza2013inferring} for interdependencies at multiple time scales, and \cite{runge2015identifying, runge2019inferring, Runge2019} for causal-discovery frameworks in spatio-temporal data. \par
This natural evolution --- from basic to sophisticated models, from {\em ad hoc} procedures to statistical rigor --- finds a natural step in implementing rigorous reassessments of inference practices. 
\citet{haas2023pitfalls} asserts that finite samples, spatial autocorrelation, and seasonal nonstationarity may inflate connectivity. Their findings suggest surrogate-based nulls, spatial block resampling, and uncertainty quantification as gold standards in statistical practices. 
Taken together, these developments move {\em networks of climate data} from heuristic discovery toward a statistically principled toolbox: Tsonis-type constructions reveal coherence, teleconnection, and tipping precursors; complex network diagnostics and causal frameworks organize these signals; modern validation closes the loop between insight and robustness. 
This provides the empirical vertex of our triad: the descriptive structures of {\em networks of} will inform the geometric realism of {\em climate data over networks} and the task-oriented machinery of {\em networks for climate data}, developed in the subsequent sections. 

\subsection{Climate Data over Networks: a Recent Story}
Data science has been a revolution in many theoretical and applied fields, and has provided many challenges within the realm of space-time data \citep[see][ for recent contributions]{anderes2020, moradi, BADDELEY2021100435, BADDELEY}. Data {\em over} networks have preoccupied both statistical and machine learning (ML throughout) communities for a longtime. Within ML community, the amount of literature is huge. However, only a small part of it is climate-focused, and mostly serves as a computational background as per Section \ref{sec-for}. Since a good part of this manuscript is related to network topologies, we mention that topological complexities through ML under the framework of data analytics are discussed in \cite{stankovic2020data}. Another relevant comment is that, in the great majority of ML related contributions, the process is assumed to be defined exclusively over the vertices of the graph. The extension to processes that are continuously defined over both graphs and edges requires substantial mathematical work, and has been challenged only recently under the paradigm of Gaussian processes over metric graphs \citep{porcu2023stationary, bolin_gaussian_2024}, which encompasses other sophisticated structures such as linear or nonlinear (generalized) networks, or over Euclidean trees \citep{anderes2020}. \par
The main advantage of working with Gaussian processes is that the dependence structured is completely specified through the covariance function \citep{porcu201930}. What makes things complicated is the covariance function typically depends on the distance between every pair of random variables defined over the graph. This is where the topology and the language of metric graphs (and metric spaces) comes into play. \par
Additionally, metric graphs might not be enough to describe sophisticated topologies. For instance, climate processes might temporally evolve over these topological structures. Building covariance functions over networks is mathematically challenging as it will be explained in subsequent sections. The task is made even more complex due to the fact that specifying the right metric is extremely challenging. Notable works in this direction have been made by \cite{ver2006spatial} and \cite{peterson2007geostatistical} to model stream river flows. \par
\cite{anderes2020} have proposed graphs with Euclidean edges, which are graphs where each edge is associated with an abstract set being in bijective correspondence with a segment of the real line. In turn, this provides each edge with a Cartesian coordinate system to measure distances between any two points on that edge. {\cite{alegria_computationally_2024} proposed efficient algorithms to simulate isotropic fields on graphs with Euclidean edges, while \cite{filosi_temporally-evolving_2025} defined temporally-evolving networks and processes on them. Finally, \cite{filosi_vector-valued_2025} extended the processes defined by \cite{anderes2020} to the multivariate setting. Different, {albeit similar,} graph structures have been considered by \cite{bolin} and \cite{bolin_gaussian_2024} through Gaussian Markov random fields. 

\subsection{Networks {\em for}: State of the Art} \label{sec-for}

A third and increasingly influential perspective concerns networks {\em for} climate data. Here,  networks represent tools, mostly {computational instruments} rather than objects of inference or observation. As previously specified, within the {\em for} class of networks the topological structure is not inferred {\em from} climate data, as in networks {\em of}, nor prescribed {\em over} a given geography, as in networks {\em over}. 
Instead, the topology is {constructed to perform a task}, which might be prediction, reconstruction, downscaling, or data assimilation. \par
We note that under {\em for}, edges are not necessarily understood as physical teleconnections. More broadly, they are operational devices where any source of information and uncertainty is used to execute a specific model objective or task. 
This implies a conceptual shift from descriptive to functional use of networks, from mapping interactions to {\em designing} them.

ML, signal processing, and statistical inference converge within this paradigm. While there is a wealth of tools especially coming from the ML literature, our manuscript will focus on Graph Signal Processing (GSP), Probabilistic Graphical Models (PGMs), and Graph Neural Networks (GNNs). The reason is, a part from their popularity, that these three classes of tools provide a hierarchy of increasing flexibility and abstraction. Further, our manuscript will argue that these tools perfectly fit inside the call 
 by \cite{ShawStevens2025} for ``disruptive computation with interpretability''. \par
We anticipate that each of these three families covers a complementary layer in the construction of networks {\em for}: GSP offers a basic (linear) and interpretable foundation layer; PGMs introduce conditional independence and uncertainty quantification; GNNs extend both to nonlinear, data-driven propagation across space and time. Some short introduction follows for each of them. \par
In GSP, the graph represents the geometry of computation and their goal is to process or reconstruct signals that live on a known graph. GSP treats a climate field as a signal defined on the nodes of a graph. Within this class of tools, spectral graph filters are typically used to smooth out or denoise, while graph Laplacians provide a natural language for interpolation and uncertainty quantification under missing observations~\citep{Shuman2013, Ortega2018}. These methods have been successful in climate modeling, in particular to design optimal sensor networks, filter satellite data, or reconstruct fields affected by observational gaps. \par 
PGMs are based on conditional dependence structures where a key role is played by the sparsity of the precision matrix, which in turn allows for scalable inference while retaining interpretability~\citep{Lauritzen1996, Friedman2008, bolin_gaussian_2024}. Dynamic Bayesian networks represent the natural generalization attained to incorporate temporal dependence, and have been largely used within the realm of causal discovery and multi-step forecasting~\citep{Shimizu2006}. Within climate science, PGMs have been used to explore directional interactions (causal teleconnections), to estimate lagged dependencies across hemispheres, and to quantify uncertainty in ensemble forecasts. \par
GNNs generalize both GSP and PGMs by learning nonlinear propagation mechanisms across the graph. They perform the taks of adapting the connectivity to optimize predictive skill~\citep{Defferrard2016}. GNNs have proved effective in basin-scale flood prediction, temperature downscaling, and urban heat mapping. 
Physics-informed variants integrate physical constraints such as mass conservation, monotonicity, and boundary conditions, aligning computational scalability with the laws of thermodynamics and hydrology. 
These methods yield improved forecasts while maintaining interpretability and physical consistency. \par
Some comments are in order. 

The objective of networks {\em for} is thus {\em functional} rather than representational, in the sense that they represent both tools for computation, but also layers within hierarchical structure, within the broad scope of solving a task as previously described. Hence, under {\em for} the abstract notion of graph is embedded into computational engine, and the same topology becomes a design variable. \par 
Networks {\em for} are a natural tool to execute the Shaw-Stevens agenda as they provide a computational framework that implements the call for disruption as per \cite{ShawStevens2025}. 
While the LSD framework remains generative and equation driven, networks {\em for} are inductive and relational.  
The coexistence of these two epistemic cultures (physical closure and relational computation) anticipates the fusion perspective developed in subsequent sections, where the triad of networks becomes a practical instrument to implement the Shaw–Stevens agenda.

\subsection{Contribution}
We guide the reader through the following facts.
\begin{itemize}
    \item We describe what we term {\em Tsonis} networks as a synonym of networks {\em of} climate data. We illustrate how to build these networks using certain classes of (dis)similarity measures. An extensive literature review opens for constructive criticism about Tsonis networks, while, at the same time, recognizing their indisputable impact on the scientific literature as well as on measures of governance for climate change. 
    \item We use the synonym {\em geophysical} networks to refer to the situation where climate data are observed over networks with geophysical or geographical constraints, that is, with a predefined topology. We illustrate three situations, namely (a) static networks (observations that are not repeated over time), (b) product spaces, where observations over a metric graph are repeated over time, but the topology of the draft does not change, and (c) dynamical metric graphs with their inducted metrics.  We put special emphasis on Gaussian processes defined over geophysical networks: we describe their covariance functions and their metrics. We discuss the impact of this part of literature on climate sciences. 
    \item We provide a thorough description of three classes of networks {\em for} while avoiding mathematical obfuscation. 
    \item A triad of networks is then followed by a triad of bridges. Bridges {\em within} provides a perspective of a \emph{corpus unicum} within the triad. For the bridges {\em between} perspective, the previous step of a bridge {\em within} becomes the perfect tool to implement the Shaw-Stevens agenda. This all culminates into a {\em meta} bridge that allows to build a new ecosystem for the coexistence of Shaw-Stevens agenda with our triad in a new sistem of knowledge that dynamically updates the data universe through data discoveries. 
\end{itemize}
The paper is written for a broad public. Hence, we have avoided mathematical formalism, while focusing on the ideas and the way things are getting done for each type of network. For more mathematically-grounded introductions, the reader can find, for each subject, a detailed reference list. 
\newpage 

\section{Networks {\em of} Data: The Tsonis Paradigm} 
\subsection{Tsonis Networks and Topologies}
Given the impact of the Tsonis school in this field of research, the remainder of the manuscript will adopt the nomenclature {\em Tsonis networks} for the described networks {\em for}. Accordingly, we shall respectively term Tsonis paradigm and Tsonis topology the framework adopted by the Tsonis school and the type of topologies described therein. As for the second, the term Tsonis topology is actually an umbrella that embraces different topologies, which are described below for the convenience of the reader. Specifically, Tsonis topologies, which are summarized in Table \ref{tab:network_types} and illustrated in Figure \ref{fig:tsonisNetworks}, include the following.
\paragraph{Regular Topologies.} The topology is very simple, as nodes are located over a lattice on the two-dimensional sphere. They exhibit deterministic patterns,  and the structure of the nearest neighbor is fixed. Examples of these networks can be found in atmosphere or ocean grids that are used for simulations (without considering teleconnections). 
\paragraph{Random Topologies.} The topology is that of graphs that are purely non-deterministic. Although there is a wealth of available models for random graphs, the literature makes abundant use of {\em Erdős–Rényi} structures, where edges are equiprobable, which implies homogeneous degree distributions. They can be used as {\em null} hypothesis within hypothesis testing statistical procedures, for instance to test for clustering and path length \citep{tsonis2006networks, donges2009complex}. Normally these networks are associated with spurious connectivity. See also the more statistical approaches by \cite{hlinka2014reliability, runge2015identifying}. 
\paragraph{Small World Topologies.} These networks exhibit low path-length and high clustering. The last indicates that neighboring nodes (that is, regions with highly correlated time series) are themselves likely to be interconnected. Short-average path length means that the network has efficient connectivity because any two nodes can be connected through a small number of edges. Under such characteristics, there is a high degree of local coherence, yet, at the same time, long-range interdependencies between regions that are very distant from each others.  Several works have shown that {\em climate is a small world Tsonis network}, and we refer to subsequent sections for more clarity and documented references.\par
\begin{table}[ht]
    \centering
    \small
    \resizebox{\textwidth}{!}{
    \begin{tabular}{|p{2.0cm}|p{3.8cm}|p{4cm}|p{3.9cm}|}
        \hline
        \textbf{Topology} & \textbf{Description} & \textbf{Climate Relevance} & \textbf{Key Studies} \\
        \hline
        Regular & Nodes on lattice, deterministic. Each node connected to fixed neighbors. & Used in climate models and spatially-regular sensor grids. & \cite{donges2009backbone},\break
        \cite{berezin2012stability} \\
        \hline
        Random & {Usually }Erdős–Rényi graphs: homogeneous edge probability. & Null-model for statistical testing. Identification of random connections. & \cite{tsonis2006networks},\break
        \cite{hlinka2014reliability},\break
        \cite{donges2009complex} \\
        \hline
        Small world & High clustering. Short (average) path length. & Captures local coherence and persistent teleconnections. & \cite{tsonis2006networks},\break
         \cite{donges2009complex},\break
         \cite{berezin2012stability},\break
         \cite{feldhoff2013geometric},\break
         \cite{ludescher2013improved},\break
         \cite{malik2012analysis},\break
         \cite{boers2019complex} \\
        \hline
    \end{tabular}
    }
    \caption{{A summary of the different types of Tsonis Networks and their uses.}}
    \label{tab:network_types}
\end{table}

Some comments are in order. Rather than being descriptive, Tsonis approach usually serves as inquisitive and diagnostic tool: the first two types of topologies are actually used to empirically test whether a climate system is a small world or not, exhibiting high clustering and short path lengths relative to a random baseline. Hence, Tsonis topologies should be taken as diagnostic archetypes --- not design families. \par 
Another relevant point is that, within the literature related with Tsonis networks, scale-free networks have covered central importance. These are topologies that exhibit power law distributions that describe the distribution of the degree associated with their nodes. In other words, most of the nodes have a small number of links, while some few have a incoherent high number of connections \citep{barabasi1999emergence}. Hence, these networks are quite different from Tsonis networks as they exhibit a high level of heterogeneity. In particular, Tsonis topologies are not cleanly power-law; further, they exhibit exponential or truncated power-law tails rather than genuine scale-free behavior; finally,
Tsonis networks are the corollary of thresholding continuous correlations, which is not a preferential attachment process as per \cite{barabasi1999emergence}.\par
Scale-free networks suggest dominant regions that have a predominant role within the climate system they belong to. For this kind of network, the probability for a given node to have $k$ links is of the order to $k^{-\gamma}$, with typically $2<\gamma<3$ \citep{barabasi2009scale}. 
Table \ref{fig:tsonisNetworks} provides a schematic comparison within Tsonis topologies. 
\begin{figure}
    \centering
    \resizebox{\textwidth}{!}{
        \begin{tabular}{ccc}
            \textbf{Regular} & \textbf{Random} & \textbf{Small World} \\
            \includegraphics[width=0.3\textwidth]{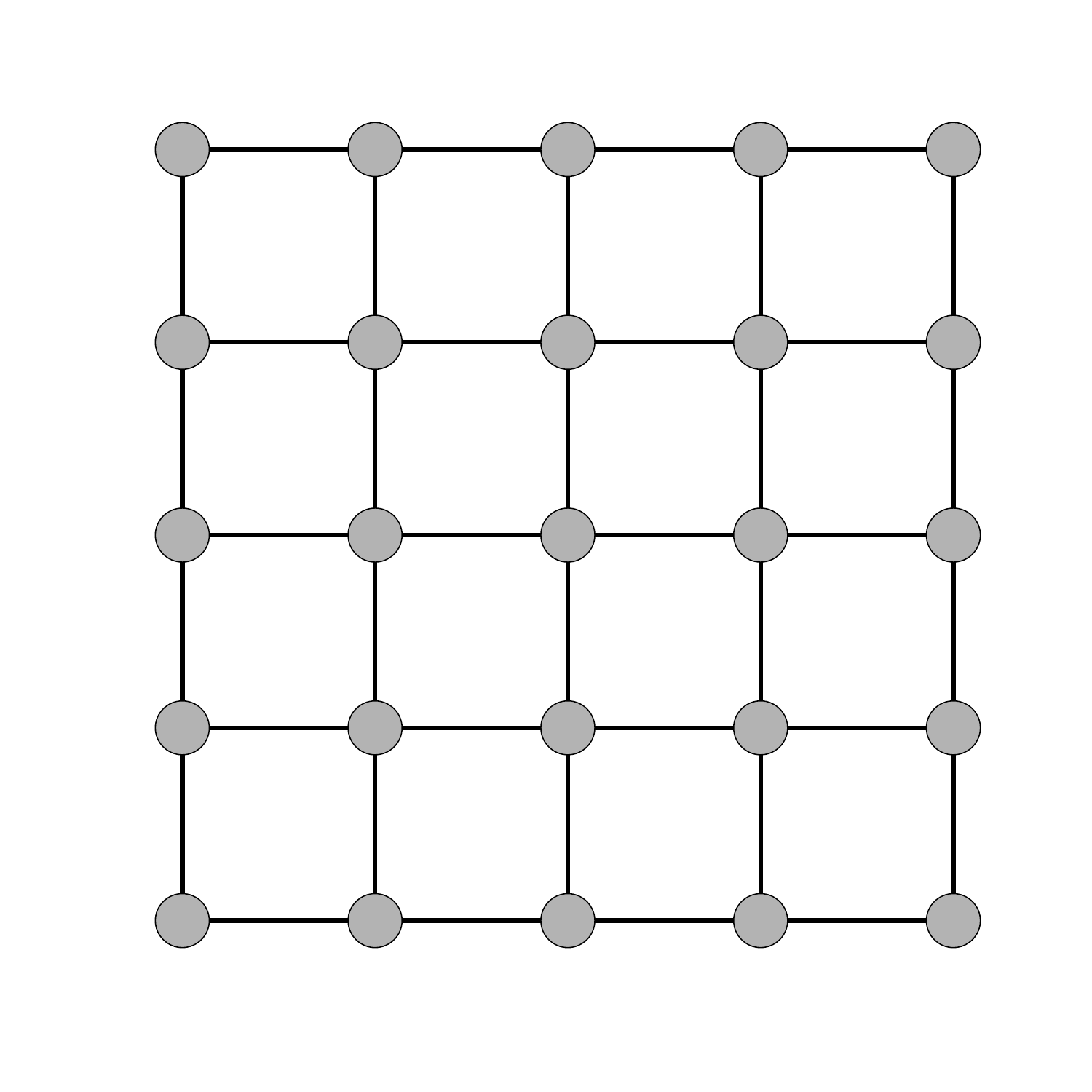} & \includegraphics[width=0.3\textwidth]{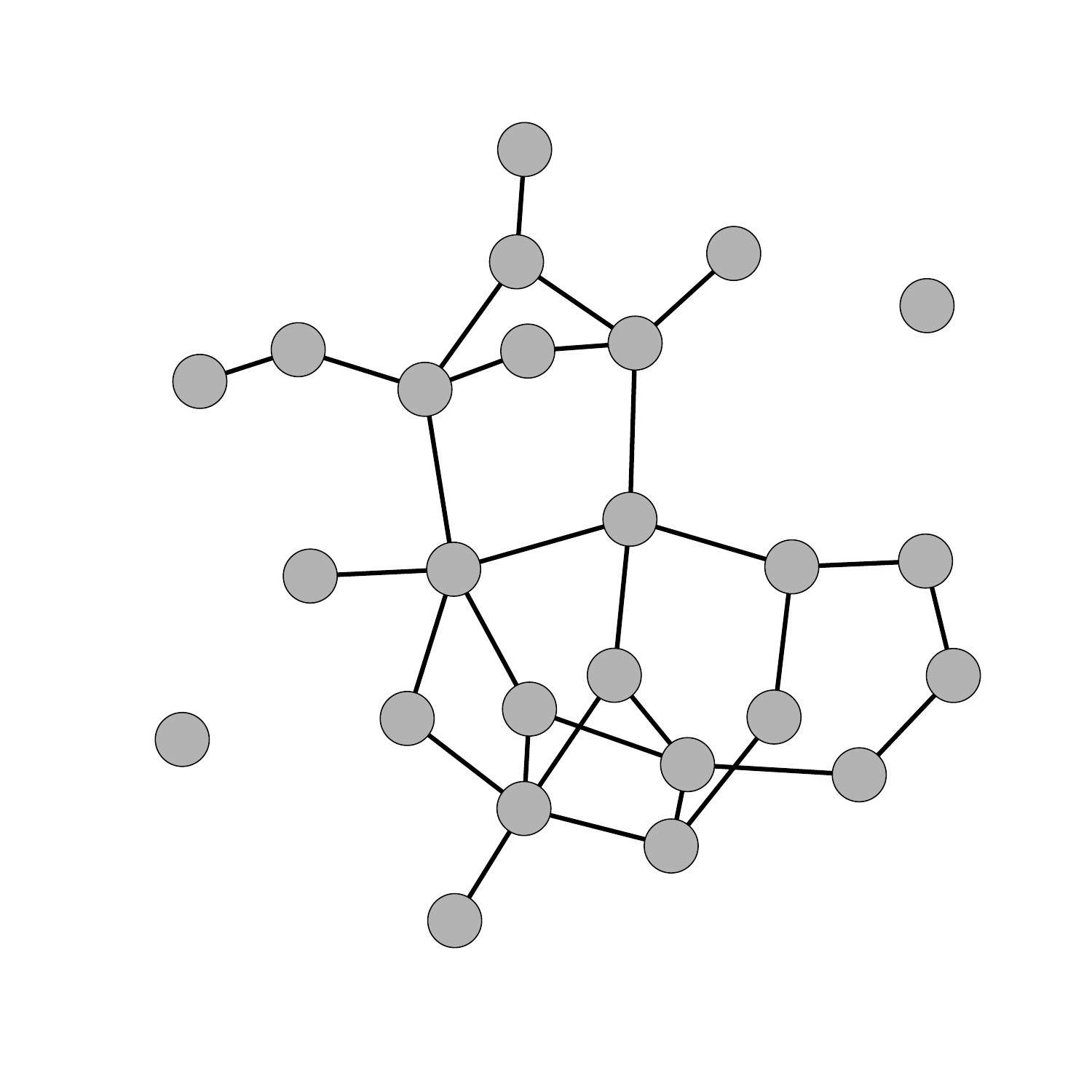} & \includegraphics[width=0.3\textwidth]{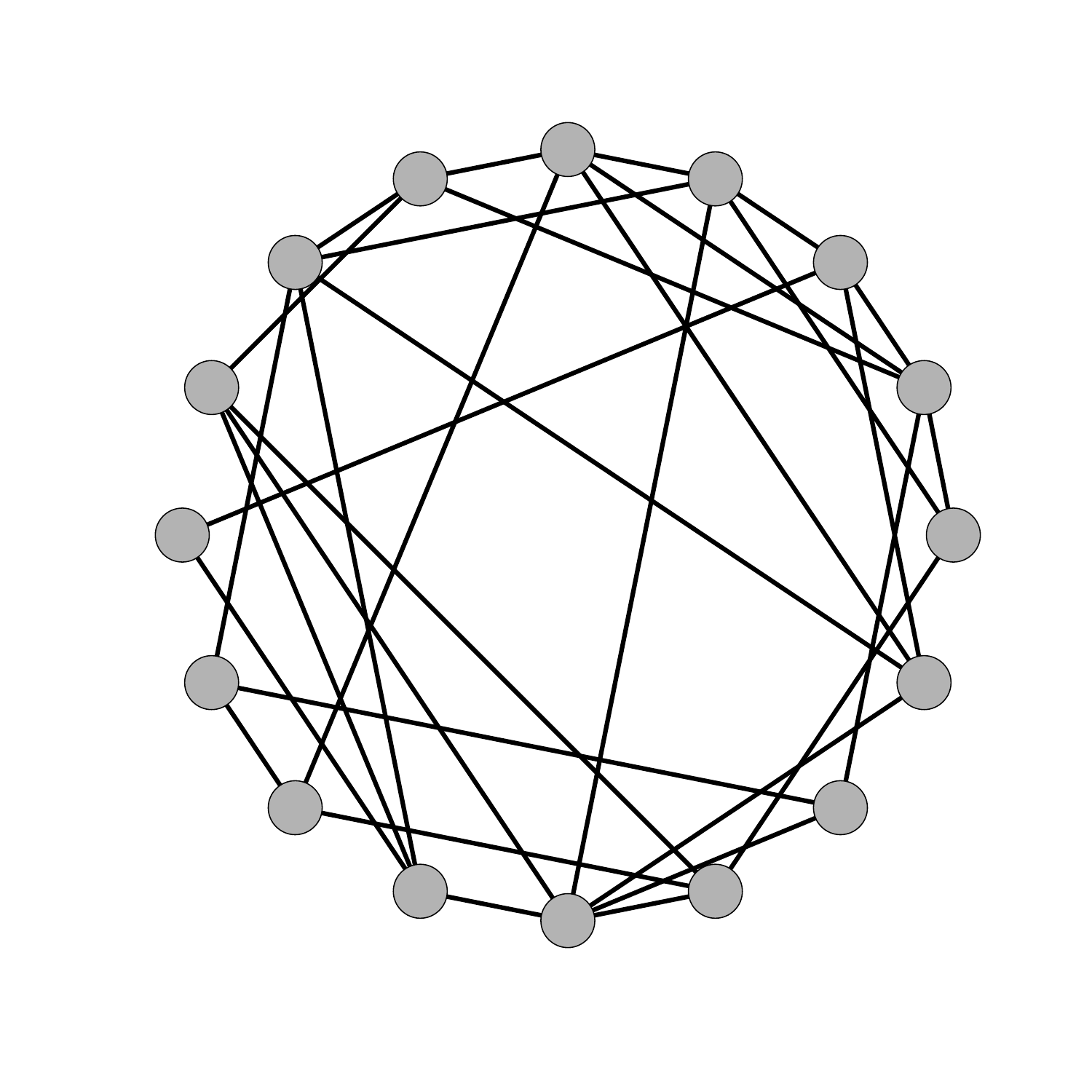} \\
            \includegraphics[width=0.3\textwidth]{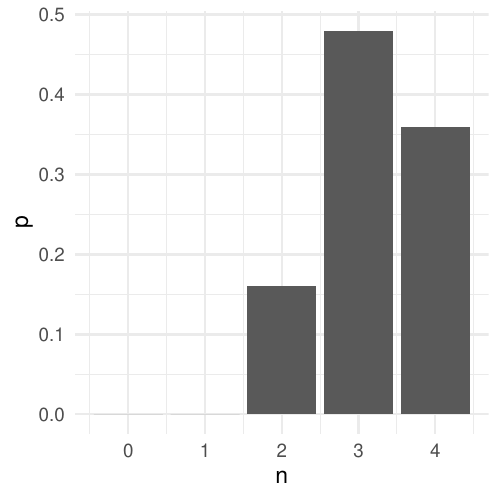} & \includegraphics[width=0.3\textwidth]{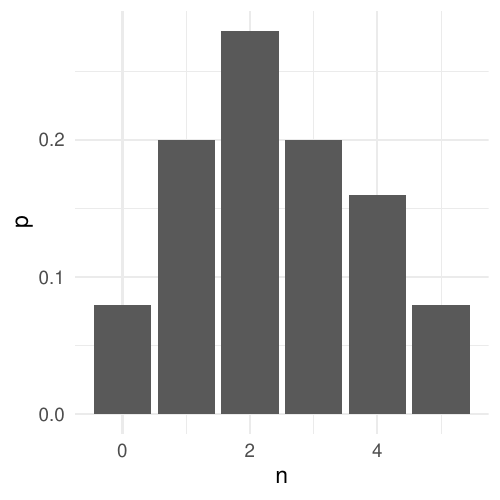} & \includegraphics[width=0.3\textwidth]{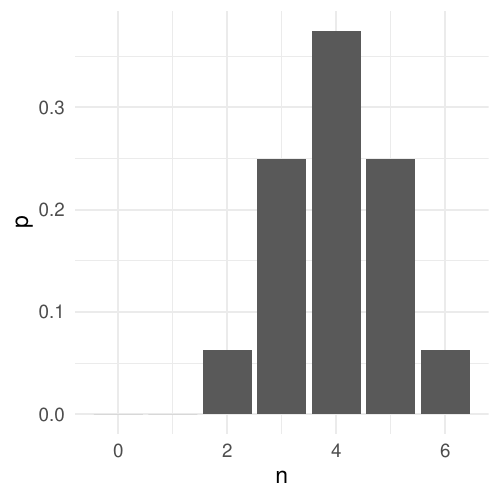} 
        \end{tabular}
    }
    \caption{A representation of different Tsonis networks with their degree distributions.}
    \label{fig:tsonisNetworks}
\end{figure}

\subsection{Tsonis Paradigm}
 To build a Tsonis network, correlations are computed between each pair of nodes based on a chosen climate variable ({\em e.g.}, surface temperature or atmospheric pressure). Then, edges are built through a simple rule: an edge between two nodes exists if and only if the correlation between them overpasses a given threshold. The resulting adjacency matrix defines a static or time-evolving climate network. \par 
To prove evidence of a small world network, \cite{tsonis2006networks} provide statistical evidence of a big discrepancy with purely random graphs in terms of cluster coefficient, as well as in terms of path length. All these evidences support the idea of a system that is not completely random, nor completely regular. \par
The structure of the network can change over time. Some nodes can keep what the authors call {\em high degree centrality} - that is, they maintain strong correlations with many other locations being even far apart across the globe. A notable example is ENSO (El Niño Southern Oscillation) in the Equatorial area. The geographical region corresponding to ENSO has a high correlation with other regions, even being far apart. This can be translated in terms of Tsonis network as the ENSO nodes have a high degree. Hence, networks help strengthen some already-established theories of teleconnections in the climate system. \par
More interesting things are actually stemming from the structure of Tsonis networks. For instance, \cite{tsonis2006networks} show that the climate network can be decomposed into clusters of nodes, which in turn are associated with teleconnection zones or climate regimes. Hence, physically meaningful subsets of the nodes can be extracted. In conclusion, Tsonis networks are extremely useful to capture local aspects (coherence) and global teleconnectivity. \par 
Special attention by this group of authors is devoted to {\em El Niño} and {\em La Niña}. During strong El Niño events, the global climate network registers notable changes in terms of connectivity patterns. \par
Tsonis network have been extremely influential for this part of the literature, as reported in the introduction to this paper. More rigor in terms of statistics is offered by \cite{donges2009complex}. Tsonis networks are the backbone for climate network approaches, such as tipping point detection and causal discovery. A tipping point is a threshold beyond which a small perturbation can lead to a qualitative, often irreversible, change in the state of a climate subsystem. Examples of tipping points include the collapse of the Atlantic Meridional Overturning Circulation (AMOC), the disintegration of polar ice sheets, and the transition from rainforest to savanna. Within the framework of Tsonis networks, tipping points can represent sharp transitions in the network topology. Such transitions can provide important indications, such as loss of coherence in the climate system (for which the sign will be a drop in connectivity), or desynchronization (seen through a sudden rise in network fragmentation), or major shifts in climate change, which can be seen through change in betweenness centrality.

\subsection{Criticism and Impact }

Tsonis paradigm is experimental rather than methodological. While its impact has been largely acknowledged in the literature, more recently some constructive criticism has emerged. The main reason is the lack of methodological and statistical rigor. The main pitfalls within Tsonis networks are associated with the following aspects, which in turn call for more rigor into future research. 
\begin{enumerate}
    \item Tsonis networks are threshold-sensitive. The topology of the network is dramatically affected by the chosen threshold. Intuitively, high threshold will allow to isolate neater clusters, while low threshold create geographic macro clusters.
    \item Autocorrelation bias happens because spatial autocorrelation can ``artificially" increase collocated correlations ({\em e.g.}, Person correlation). 
    \item Sampling variability is a big issue and has been examined to a very limited extent.
\end{enumerate}
\cite{haas2023pitfalls} is a notable example of methodological innovation and statistical rigor inside the Tsonis architecture. 
They provide constructive criticism on the choice of similarity measures and their impact. They prove that there is a risk of having spurious links because of finite sample effects, spatial autocorrelation, and seasonal nonstationarity. The alternative proposal by \cite{haas2023pitfalls} is that of surrogate-based null models in concert with bootstrapping techniques in order to understand the {\em significance} of inferred edges. Further, they claim to be able, through this approach,  to quantify uncertainty in network metrics. \par
\cite{haas2023pitfalls} point the ways to Tsonis networks improvement, through, for instance:
\begin{enumerate}
    \item objective criteria to threshold selection;
    \item robustness for the selection of links;
    \item improvement or extensions such as dynamical and multivariate.
\end{enumerate}
Hence, \cite{haas2023pitfalls} contribution can be seen as a rigorous statistical formalism that allows for improvements within the realm of Tsonis networks. The main innovations in \cite{haas2023pitfalls} come from spatial block-bootstrap methods as a replacement of threshold selection. Further, \cite{haas2023pitfalls} implicitly suggest an integration of data-driven with model-driven choices in climate sciences.  \par 
The scientific impact of Tsonis paradigm is unquestionable: Tsonis frameworks created an insatiable thread of contributions, where many of them can be seen as methodological innovation, or as more sophisticated architectures that use Tsonis topology as a baseline. Within the extension of similarity measures, mutual information and nonlinear metrics have been introduced by \cite{donges2009complex}, and subsequently by \cite{berezin2012stability} and \cite{deza2013inferring}. Rigorous approaches to synchronization have been taken by \cite{boers2019complex} and by \cite{stolbova2014topology}. Early warnings (tipping points) have been challenged in \cite{tsonis2008topology} and further studied by \cite{berezin2012stability}. Then, \cite{runge2015identifying} supported causality-connection findings through graphical models.   \cite{wiedermann2016climate} worked on ENSO predictability. Similar works can be found in \cite{malik2012analysis}. In causal inference, notable works are those of \cite{donges2015unified} and \cite{runge2019inferring}. While the previously-mentioned contributions might be seen as generalizations of a basic structure, it is also true that Tsonis paradigm triggered a diversified portfolio of methodological innovations that range from statistics, dynamical systems and causality, and that intersect with networks {\em for} in several directions (for instance, GNNs). 
Applications have clearly benefited of these methodological insights. To mention a few of them, early detection of tipping points (for instance in monsoon systems) can be found in \cite{Boers2021, stolbova2014topology}.  Circulation regimes in the arctic region can be found in \cite{Boers2021, feldhoff2013geometric}. Analysis of extreme events (floods) are accounted \cite{zhou2022hydrograph}.}  ENSO, NAO, IOD studies can be found in \cite{stolbova2014topology}; there are also studies of extreme rainfall patterns \citep{boers2019complex} and Antarctic/Arctic dynamics \citep{Boers2021}.

\subsection{Tsonis Networks and Climate Governance}
Within the past two decades, there has been a clear shift towards data-driven decisions in climate governance. The influence of Tsonis networks extends beyond academia into climate resilience planning, early-warning design, city governance, and multi-actor policy coordination. \par
Regarding early warning systems, all agencies at both national and international level, such as NOAA, ECMWF, and WMO, have accepted metrics based on networks to integrate geophysical indicators \citep{runge2019inferring}. See for instance how \cite{Boers2021} has established important transitions in terms of abrupt ENSO phase shifts, and this enables anticipation of regional drought or flood regimes. The same remark applies for national agencies in India and East Africa with NGOs such as Red Cross Climate Center. \par
Networks representations are definitely helpful for urban planning. The celebrated projects under the \textit{C40 Cities} and \textit{Cities for Climate Protection} (CCP) initiatives use graphs and network-based methodologies to identify cities vulnerabilities and cross-regional spillovers \citep{Acuto2020, Abel2025}. \par
Tsonis networks can be seen as \textit{boundary objects} \citep{Star1989}: real or abstract entities that can be used across heterogeneous communities while keeping their identity. Applying this concept to climate governance, we see that Tsonis networks are simultaneously a scientific paradigm, an instrument for data science visualization, and an instrument for policies. \par
The logic of Tsonis networks aligns with the post-Paris logic of polycentric governance. National contributions are part of a network commitment, and agencies are now calling for network interdependence \citep{UNFCCC2024}. The \textit{Climate Technology Centre and Network} (CTCN), a UNFCCC entity, supports the use of climate networks in guiding adaptation investment, especially in Least Developed Countries (LDCs) \citep{CTCN2022}. Likewise, World Bank resilience planning now includes network fragility metrics in urban diagnostics. \par
In conclusion, Tsonis networks go beyond data science and enter the realm of decision science, influencing a wealth of directions where important decisions need to be taken for a better future of our planet. 

\section{Climate Data {\em over} Networks}
\subsection{What are geophysical Networks? }
The previous sections have focused on networks of climate data as topologies that shape up from statistical dependencies, such as correlations, information-theoretic links, or any other form of statistical dependencies between space-time climate variables. These networks are typically inferred from data, with the goal of revealing latent connectivity patterns in the climate system. \par
The next paradigm considers climate data {\em over} networks: here, a geophysical processes is modeled on top of pre-specified spatial or structural graphs. These networks may arise from hydrological catchments, transportation grids, sensor layouts, or river basins, and are typically defined \emph{a priori} based on physical, infrastructural, or geometric constraints.\par
Throughout, the predefined networks described in this section will be termed geophysical networks. Accordingly, the paradigm ruling geophysical networks will be termed a geophysical paradigm. \par
The language of metric graphs is the best platform to mathematically describe networks of the geophysical type. Recent research has proven how these graphs are extremely flexible in that they allow for space-time processes as well for spatially dynamical processes, where both vertices (nodes) and edges can appear or disappear over time.  \par
A very important innovation with geophysical graphs is that the great majority of earlier contributions considers processes that are exclusively defined over vertices. The definition of a process that is continuously indexed over both vertices and edges requires a fair amount of mathematical work. Geophysical graphs are ingenious topological structures that allow to generalize linear networks to nonlinear edges. Further, the process defined over such structures can have realizations over any point over the edges, and not only in the nodes.  \par 
More simplified forms of networks have already received attention within the ML community \citep{alsheikh2014machine,  georgopoulos2014distributed, hamilton2017representation, pinder2021gaussian, borovitskiy2022isotropic}. We already mentioned the increasing availability of this kind of networks for spatial data \citep{cressie2006spatial, gardner2003predicting, ver2006spatial, peterson2013modelling, peterson2007geostatistical, montembeault2012impact} and point processes \citep{xiao2017modeling, perry2013point, deng2014ginibre, baddeley2017stationary}. \par
Surprisingly, all the above mentioned contributions refer to static networks \citep[see also the more recent approach by][]{bolin_gaussian_2024}, where temporal dynamics have not been part of the game, till very recently.  \par
\cite{porcu2023stationary} and \cite{tang_space-time_2024} overcome this problem by considering product metric spaces, where one space is the network and the other is time. However, a limit of this construction is that the topology cannot be adapted to change over time. This drawback has attracted the interest of several applied scientists \citep{hanneke_discrete_2010}, but produced very limited results. See fore instance \cite{mankad_structural_2013, cherifi_community_2019, lim_link_2019, divakaran_temporal_2020, rossi_modeling_2013} within the framework of structural changes detection. All these approaches do not allow for processes that are continuously defined over the network. \par 
\cite{filosi_temporally-evolving_2025} proposed time-evolving graphs where (discrete) time is either linear o periodic. They argue that such a choice fits several real-life situations, to include for long-term as well as periodic behaviors. They provide the example of temperatures in a given geographical area, where there might be strong correlations between: (i) contiguous spatial points at a given time (represented by means of spatial edges); (ii) the same points considered at contiguous times (represented by means of temporal edges between temporal layers) and (iii) the same points considered at the same periods of the year, which are considered in the model as they are exactly the same point in the temporally evolving graph. \par
A gentle mathematical description of geophysical networks follows. When defining geophysical networks, we are referring to, either, (a) a static metric graph, (b) a product space involving metric graphs, or again (c) a temporally dynamical metric graph. A brief illustration about metric graphs follows. A closer focus will be given to a broad class of metric graphs that are termed graphs {\em with Euclidean edges} after \cite{anderes2020}. \par
\begin{figure}
    \centering
    \includegraphics[width=0.8\textwidth]{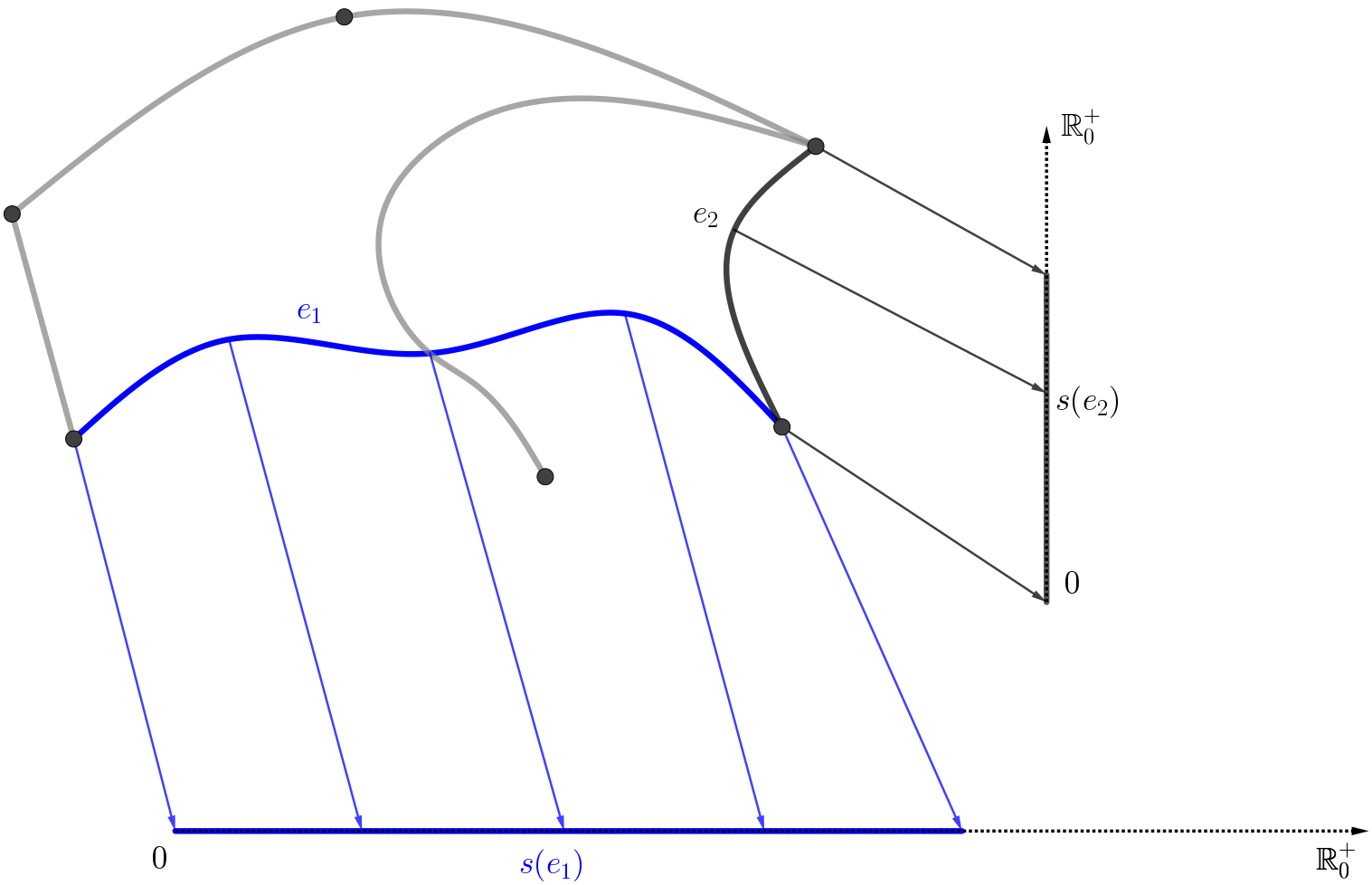}
    \caption{A graph with Euclidean edges, where the bijections between the edges $e_1$ and $e_2$ and their abstract segments $s(e_1)$ and $s(e_2)$ have been highlighted. Figure adapted from \cite{filosi_temporally-evolving_2025}.}
    \label{fig:graphEE_illustration}
\end{figure}

A metric graph is a graph whose edges are not only relations between vertices, but also sets of \emph{physical} points. More formally, a graph $\mathcal{G}=(V,E)$, where $V$ is the set of vertices and $E\subset V\times V$ is the set of edges, becomes a metric graph when each edge $e\in E$ is associated to a geometric segment, say $s(e)$, of the real line (see Figure \ref{fig:graphEE_illustration}). In such a way, each point on the segment $s(e)$ has its own geometrical meaning. Typically, the \emph{length} of the segment $s(e)$ coincides with the \emph{geometrical length} of the edge $e$, which is usually inherited from the meaning of the graph $\mathcal{G}$. For example, if $\mathcal G$ represents a road network, where nodes are crossroads and edges are connections among them, the length of an edge $e$ could be naturally chosen as the physical length of the relative road. However, there are no constraints that force this choice. The advantages of such a topological structure is that it is possible to define distances among any couple of physical points.\par
Once such a structure has been defined, it is natural to consider some extension: for instance, it is possible to define a product space between a metric graph and an arbitrary domain. Perhaps the most intuitive example is to consider a graph with Euclidean edges cross time \citep{porcu2023stationary}. This means that it is possible to consider every point in the graph at different time instants. Thinking about road networks, this allows to model a city over an arbitrary time set, such as a day or a year (see Subsection \ref{sssec:EmilioSpatioTemporal}). This approach paves the way to the extremely reasonable question: may graphs change over time? \cite{filosi_temporally-evolving_2025} have defined an ad-hoc mathematical structure that represents both graph with an arbitrary dynamics over discrete time (that is: nodes and edges may appear or disappear at subsequent time instants) or with periodic evolution (see Figure \ref{fig:periodicConstruction}). Although this may seem a strong hypothesis, they argue that it may appear in some specific context, such as river deltas governed by the tide or road networks where traffic lights induce a periodic dynamic of the flow. \par 

A linear network is a special case of a metric graph that is basically the union of nodes through linear segments on the plane. For such a case, measuring distances over a linear network is straightforward. However, defining distances (called metrics) over a general metric graph is challenging. \par
\begin{figure}[ht]
    \small
    \centering
    \begin{tabular}{>{\centering\arraybackslash}m{3cm}>{\centering\arraybackslash}m{6.5cm}>{\centering\arraybackslash}m{3.5cm}}
        \begin{circuitikz}[thick, 
        scale=1.1]      
            \coordinate (A) at (1,0);
            \coordinate (B) at (2,0);
            \coordinate (C) at (2,2);
            \coordinate (D) at (2,4);
            \coordinate (E) at (3,0);
            \coordinate (F) at (3,2);
            \coordinate (G) at (3,4);
            
            \draw (A) to[short, l_=$5$, *-*] (B);
            \draw (B) to[short, l_=$10$, name=r2, *-*] (C) to[short, l_=$10$, name=r2, *-*] (D);
            \draw (E) to[short, l_=$10$, name=r2, *-*] (F) to[short, l_=$10$, name=r2, *-*] (G);
            \draw (B) to[short, l_=$5$, name=r3, *-*] (E);
            \draw (C) to[short, l_=$5$, name=r3, *-*] (F);
            \draw (D) to[short, l_=$5$, name=r3, *-*] (G);
            
            \node[above left] at (A) {A};
            \node[above left] at (B) {B};
            \node[above left] at (C) {C};
            \node[above left] at (D) {D};
            \node[above left] at (E) {E};
            \node[above left] at (F) {F};
            \node[above left] at (G) {G};
        \end{circuitikz} & 
        \begin{circuitikz}[thick, 
        scale=1.1]      
                \coordinate (A) at (0,0);
                \coordinate (B) at (2,0);
                \coordinate (C) at (2,2);
                \coordinate (D) at (2,4);
                \coordinate (E) at (4,0);
                \coordinate (F) at (4,2);
                \coordinate (G) at (4,4);
                
                \draw (A) to[resistor, l_=$5\Omega$, *-*] (B);
                \draw (B) to[resistor, l_=$10\Omega$, name=r2, *-*] (C) to[resistor, l_=$10\Omega$, name=r2, *-*] (D);
                \draw (E) to[resistor, l_=$10\Omega$, name=r2, *-*] (F) to[resistor, l_=$10\Omega$, name=r2, *-*] (G);
                \draw (B) to[resistor, l_=$5\Omega$, name=r3, *-*] (E);
                \draw (C) to[resistor, l_=$5\Omega$, name=r3, *-*] (F);
                \draw (D) to[resistor, l_=$5\Omega$, name=r3, *-*] (G);
                
                \node[above left] at (A) {A};
                \node[above left] at (B) {B};
                \node[above left] at (C) {C};
                \node[above left] at (D) {D};
                \node[above left] at (E) {E};
                \node[above left] at (F) {F};
                \node[above left] at (G) {G};
        \end{circuitikz} & \begin{tabular}{|c|c|c|c|}
            \hline
            $N_1$&$N_2$&$d_{SP}$&$d_R$ \\
            \hline
            $A$&$B$&$5$&$5$ \\
            $B$&$G$&$25$&$12.1$ \\
            $E$&$G$&$20$&$12$ \\
            $C$&$F$&$5$&$3.6$ \\
            \hline
        \end{tabular}
    \end{tabular}
    \caption{A physical network (left), its associated resistor graph (center) and distances between some couple of nodes (right). Distances are always measured in meters. $d_{SP}$ denotes the shortest-path distance, while $d_R$ is the resistance distance.}
    \label{fig:resistanceGraph}
\end{figure}
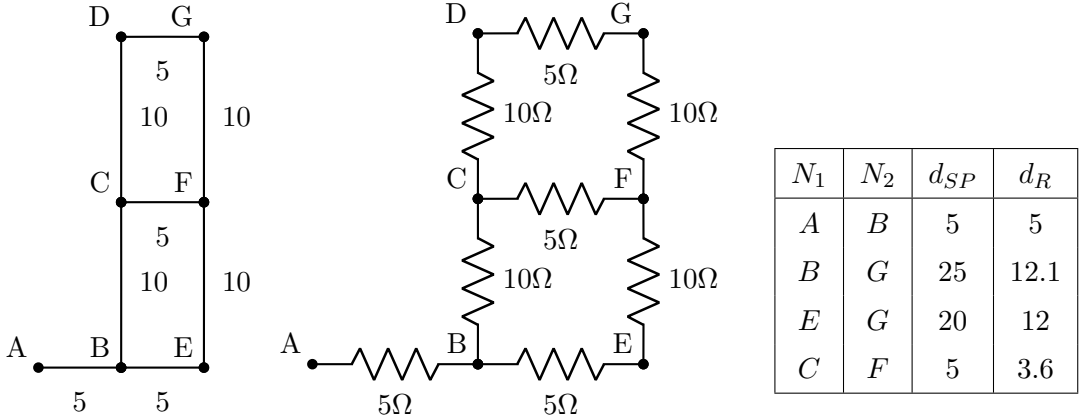

One intuitive way to define distance between any two points over the metric graph is by using the shortest-path distance, which measures the total length of the shortest path that connects two nodes. This metric has been proved by \cite{anderes2020} to have limited applicability (details coming later). As an alternative, they propose the effective resistance distance, which takes into account \emph{all} the paths that connect two given nodes. As a consequence, the resistance metric can be more useful than the shortest-path in several contexts. To give an example, think about a pedestrian network as depicted in Figure \ref{fig:resistanceGraph}: people that have to go from point $B$ to point $G$ have 3 possible paths, all with the same total physical length of $25m$. Therefore, it may be reasonable to ``reduce'' the distance between them, as they are more interconnected then the nodes $B$ and $A$ (for which there is just one path).  Effective resistance distance overcomes this problem as it models a resistor graph, where edges are electrical resistors (whose resistances are proportional to edges lengths) and nodes are junction points. Thus, for example, two resistors connecting two nodes in parallel result in a lower effective resistance, according to the Ohm's law, whilst two edges connected by a single path will have a resistance distance equal to the shortest-path metric. For a more detailed analysis of the effective resistance distance, the reader is deferred to, for instance, \cite{jorgensen_hilbert_2010}. \cite{anderes2020} found an ingenious method for extending the classical resistance distance to geophysical networks, where distances can be computed not only between nodes, but also between any pair of points on the graph.

\subsection{Climate Processes over Geophysical Networks}

A random variable is, roughly speaking, a mathematical object that accounts for some random behavior. Such objects are extensively used to rigorously model, predict and asses the variability of real random phenomena, including climate. As an example, it is of obvious interest to know what the temperature will be tomorrow in a given spatial point. \par
A \textit{random field} is a collection of random variables, that is: a set of random variables indexed by some additional (nonrandom) variable. This can be formally represented via $\{Z(\boldsymbol{x})\,:\,\boldsymbol{x} \in X\}$, being $X$ the above-mentioned set of indices. Thinking about the temperature example, a random field could be the temperature at a given spatial point through the hours of a day: in such a case the formal representation could be $\{Z(t)\,:\,t=0,1,\dots,23\}$, where $t$ is interpreted as the time instant ($Z(t=0)$ indicates the temperature at 00:00, whilst $Z(t=7)$ is the temperature at 07:00 and so on). Another example may be the total precipitation amount (measured in mm) on a large geographical region in a given day. In this case the index set $X$ is the set of spatial locations belonging to the region and $Z(\boldsymbol{x})$ is the local precipitation amount at the specific location $\boldsymbol{x}\in X$. In many real-world climate applications, we are interested not only in either spatial or temporal dynamics, but in spatio-temporal one. Think about at the precipitation amount in a large region through one year period. In such a case, the set $X$ could be rigorously defined as the cartesian product of a space domain (e.g. $\mathcal S$) and a temporal one (e.g. $\mathcal T$). As a consequence, the random field $Z$ will be $\{Z(\boldsymbol{x},t)\,:\,\boldsymbol{x} \in \mathcal S,\,t \in \mathcal T\}$.  \par
\begin{figure}[ht]
    \centering
    \resizebox{\textwidth}{!}{
        \begin{tabular}{m{0.43\textwidth}m{0.53\textwidth}}
            \includegraphics[width = 0.42\textwidth]{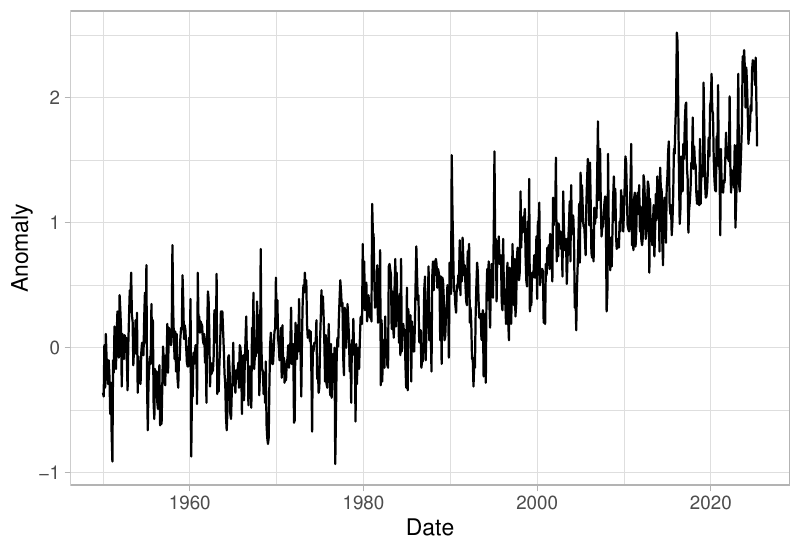} & \includegraphics[width = 0.51\textwidth]{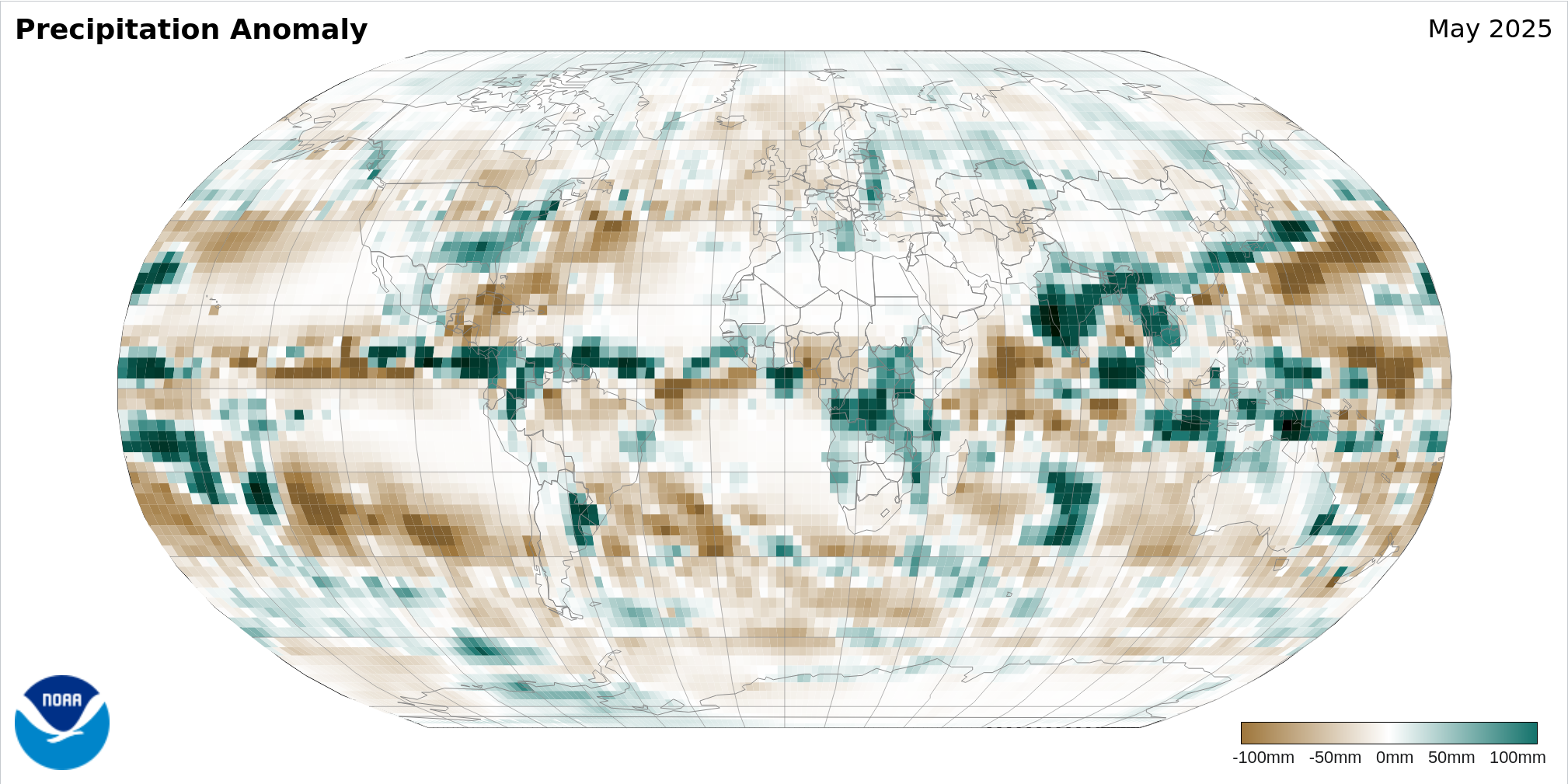}
        \end{tabular}
    }
    \caption{Left: monthly global land temperature anomalies since 1950. Right: precipitation anomalies in May 2025 on a $2.5^\circ \times 2.5^\circ$ grid. Data for the first figure and the second figure has been taken from \cite{NOAA_climate_data}.}
    \label{fig:globalLandTemperatures}
\end{figure}

To help the reader understanding the above-introduced concepts, Figure \ref{fig:globalLandTemperatures} shows two climate phenomena that may be seen as random fields. More specifically, Figure \ref{fig:globalLandTemperatures} (left) shows the time series of global land temperature anomalies over the period 1950-2024. In this case, $\mathcal{T}=\{(y,m), 1950 \leq y \leq 2024, 01 \leq m \leq 12 \}$, being $y$ the year and $m$ the month. Figure \ref{fig:globalLandTemperatures} (right) shows precipitation anomalies in May 2025 on a regular grid of $2.5^\circ \times 2.5^\circ$. In this case, the spatial domain is the set $\mathcal{S}=\{(lat, long)\,:\,lat \in \{88.75^\circ N, \dots , 88.75^\circ S\}, long \in \{\{88.75^\circ E, \dots , 88.75^\circ W\}\}\}$. Although, in some sense, both the phenomena are deterministic, the chaotic behavior of climate and whether, together with our incomplete knowledge of the system, force us to cope with some \emph{uncertainty}: here the just exposed machinery definitely help. Indeed, we \emph{model} climate and weather phenomena as random fields, to have a rigorous framework where we can measure and handle uncertainty.

\subsection{Covariance Functions}
A random field is called Gaussian when, for any arbitrary collection of points in the reference domain, say $\boldsymbol{x}_1,\ldots,\boldsymbol{x}_N$, with $N$ arbitrary, the random vector $(Z(\boldsymbol{x}_1),\ldots,Z(\boldsymbol{x}_N))^{\top}$ has a multivariate Gaussian distribution. Here, $\top$ is the transpose operator. Gaussian random fields are completely described by their covariance function, which is a linear measure of dependence between the random variable $Z$ at some point $\boldsymbol{x}_1$ (or at the space-time coordinate $(\boldsymbol{x}_1,t_1)$) and the random variable $Z$ at some point $\boldsymbol{x}_2$ (or similarly, at the space-time point $(\boldsymbol{x}_2,t_2)$). The covariance function describes linear concordance or discordance. That is, if the two variables tend to increase or decrease together (concordance), then they will show positive covariances, whilst if one will usually decrease when the other grows and vice versa, they will show a negative covariance (discordance). Finally, if two random variables are completely independent, that is: the behavior of one gives absolutely no information about the other, then they will have covariance identically equal to zero. The converse is true if the random field is Gaussian. Think about the mean temperature during one day in two close geographical points: then it is reasonable to assume that, but for some specific local meteorological phenomenon, an increase in the temperature in the first point corresponds to an increase in the second point as well. Conversely, if we think at precipitation and temperature in the same point, it is likely that an increase in one corresponds to a decrease for the other. \par
That said, when we move from random variables into random fields, the covariance can be computed between any two points of the domain and, as a consequence, becomes a function. More precisely, assuming that our domain is a metric graph $\mathcal G$ times a temporal set $\mathcal{T}$, the covariance function of a process $Z:\mathcal G \times \mathcal T \to \mathbb R$ is defined via:
\begin{equation*}
    C_Z((\boldsymbol{x},t), (\boldsymbol{x}',t')):=\text{cov}\left(Z(\boldsymbol{x},t), Z(\boldsymbol{x}',t') \right),
\end{equation*}
where $(\boldsymbol{x},t)$ and $(\boldsymbol{x'},t')$ belong to $\mathcal{G} \times \mathcal{T}$. It is mandatory to note that covariance functions need to satisfy some conditions: they must be symmetric (that is: the covariance between $A$ and $B$ is the same as between $B$ and $A$) and positive definite. This last condition is usually particularly challenging \citep{berg2008stieltjes}. However, there are some standard methods to build valid covariance functions. For instance, given a spatial covariance function $C_\mathcal{S}$ on a domain $\mathcal{S}$ and a temporal one $C_\mathcal{T}$ on $\mathcal{T}$, then the function $C_{\mathcal{S}\times \mathcal{T}}((\boldsymbol{x},t), (\boldsymbol{x}',t')):=C_\mathcal{S}(\boldsymbol{x}, \boldsymbol{x}')\, C_\mathcal{T}(t, t')$ is a valid covariance function on $\mathcal{S} \times \mathcal{T}$ (and is called \emph{separable} covariance function). Although the separability approach is very flexible, as it is possible to combine a wealth of purely spatial and purely temporal covariance functions, several studies \citep[see][with the references therein]{porcu201930} have argued that it cannot cope with the majority of spatio-temporal complex dynamics. Some work in this direction has been recently done, and next we report the main ideas.

\subsubsection{River networks}
\begin{figure}[ht]
        \centering
        \resizebox{\textwidth}{!}{
        \begin{tabular}{cc}
            \includegraphics[width=0.6\linewidth]{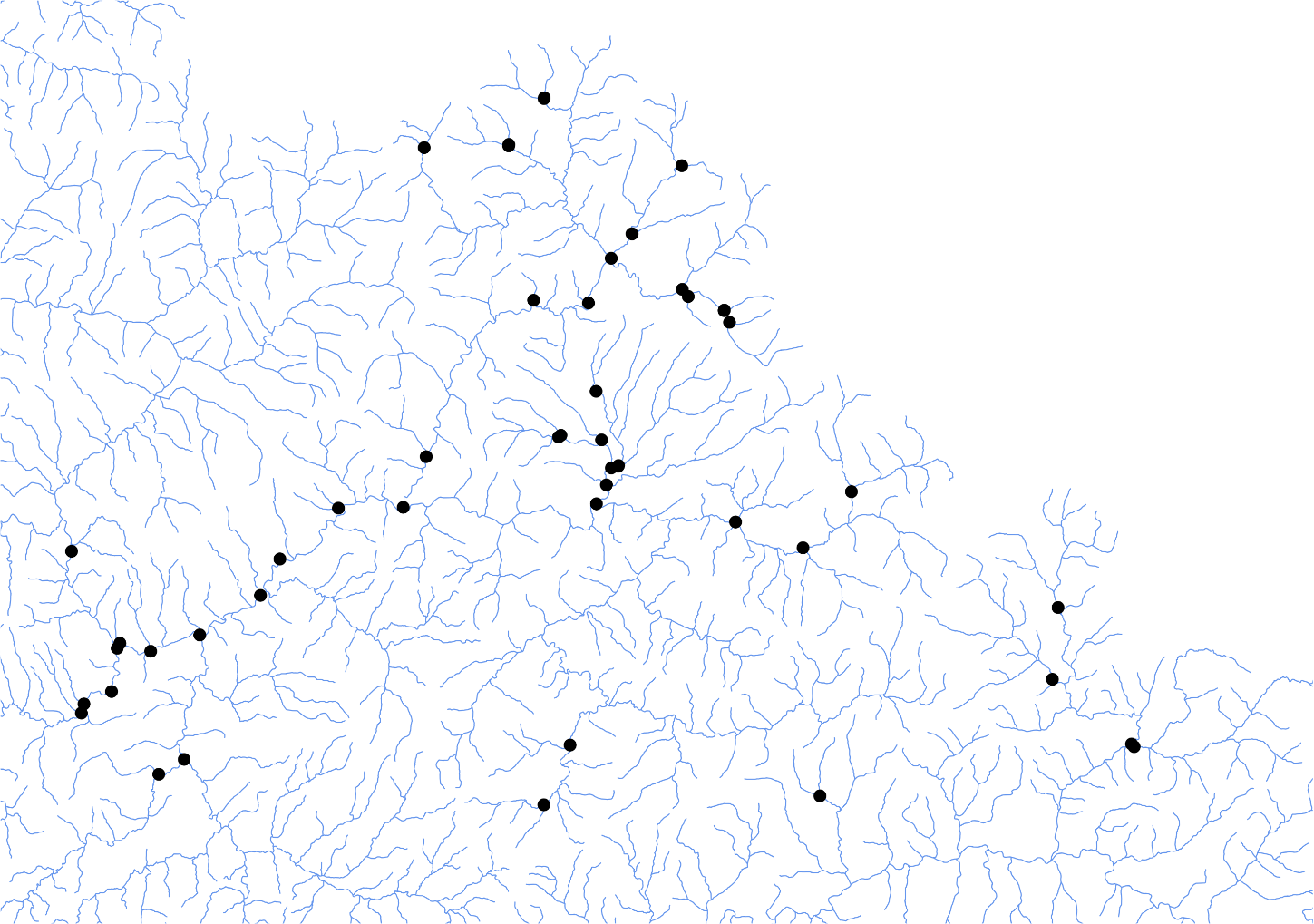} & \begin{tikzpicture}[
            thick, 
            every node/.style = {
                draw, circle, minimum size=2mm, inner sep=0pt, fill=black
            }, 
            ->-/.style = {
                decoration={
                    markings, mark=between positions 0.55 and 0.75 step 0.5 with {\arrow{Triangle[scale=1.4]}}
                },
                postaction={decorate},
                thick, draw=black
            },
            scale=0.8
        ] 
            \node (A1) at (0,0) {};
            \node (A2) at (2,0) {};
            \node (A3) at (4,0) {};
            \node (A4) at (5,0) {};
            \node (A5) at (3,0) {};
            \node (A_1) at (6,-2) {};
            \node (B1) at (1,-4) {};
            \node (B3) at (4,-4) {};
            \node (B4) at (6,-4) {};
            \node (B_1) at (1,-6) {};
            \node (C1) at (3,-8) {};
            
            \draw [->-] (A1) -- (B1);
            \draw [->-] (A2) -- (B1);
            \draw [->-] (A3) -- (B3);
            \draw [->-] (A4) -- (A_1);
            \draw [->-] (A5) -- (C1);
            \draw [->-] (A_1) -- (B4);
            \draw [->-] (B1) -- (B_1);
            \draw [->-] (B3) -- (C1);
            \draw [->-] (B4) -- (C1);
            \draw [->-] (B_1) -- (C1);
        \end{tikzpicture}  
        \end{tabular}
        }
        \caption{Left: Clearwater River Basin in Idaho (USA) and $50$ points of interest on it. Right: how a river basin can be modeled through an (oriented) tree.}
        \label{fig:riverBasin}
    \end{figure}
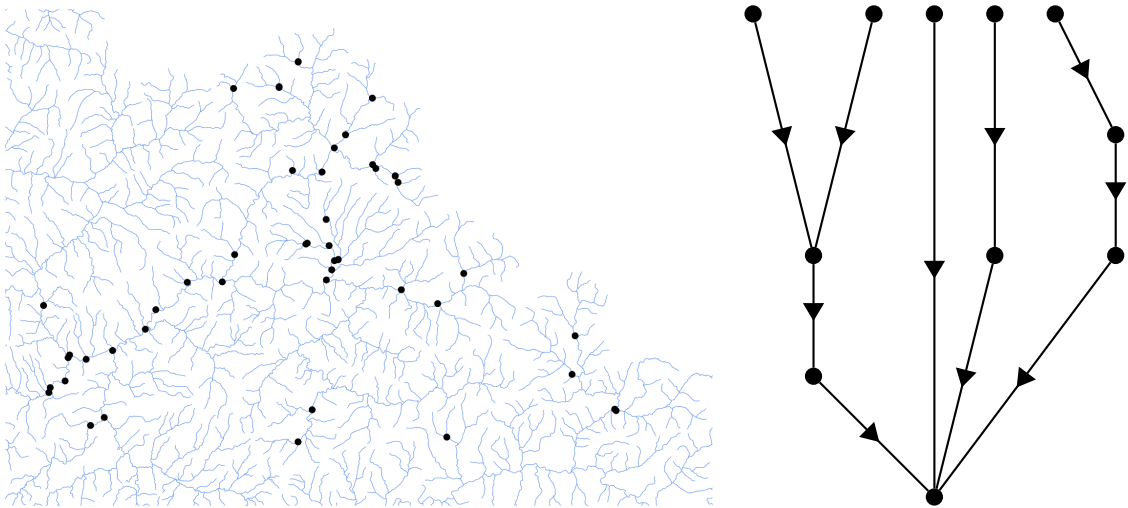

    Perhaps the first and most intuitive example of a geophysical network is a river basin, that is: the network formed by a river, its tributaries, their tributaries and so on up to the springs. This network is, usually, a tree, meaning that given two arbitrary points on the network there exists exactly one path connecting them. This is the case since usually a stream merges with others, and never splits itself in two. Figure \ref{fig:riverBasin} shows an example of such a river basin \citep[Figure 4]{porcu2023stationary} and how it is possible to rigorously model it. 
    \cite{ver2006spatial} have been the first to properly define valid covariance functions based on the stream distance, as several previous analyses have been performed using spherical or Euclidean distance, leading to possible non-positive defined covariance functions or unrealistic models. A challenge in modeling river networks is that water \emph{flows}, that is: two close points could share either the same water or some water, or even they could have independent streams. This makes models based on shortest-path distance unrealistic in several cases. Think, for instance, at the estimation of the concentration of some chemical molecule through the basin: the concentration of two streams that merge into another after even a short distance have \emph{a priori} no reason to be correlated. However, the concentration of the stream below the merging point should be the (weighted by volume) mean of the concentrations of the two original streams. For this reason, \cite{ver2006spatial} have defined an ingenious framework that result in null covariance for points that are not \emph{flow-connected}, that is that do not share water, and positive covariance for flow-connected points.\par
    {Their method relies on the careful integration of proper processes defined on the river network: they \emph{define} a process in a given point $\boldsymbol{x}$ as the integral of a (weighted by water volume and by distance) white noise over the set of all points whose water flows eventually through $\boldsymbol{x}$. This allows to achieve the desired result of independence on points that are not flow-connected and ends up in a quite simple, albeit perhaps a bit ugly, covariance function:
    \begin{equation*}
        C_\mathcal{S}(\boldsymbol{x}_1, \boldsymbol{x}_2)=\begin{cases}
            0 \qquad &\text{if $\boldsymbol{x}_1$ and $\boldsymbol{x}_2$ are not flow-connected},\\
            C_1(0) + \nu_{\boldsymbol{x}_1}^2 \qquad &\text{if} \boldsymbol{x}_1=\boldsymbol{x}_2,\\
            \prod_{k \in B_{\boldsymbol{x}_1, \boldsymbol{x}_2}} \sqrt{\omega_k} C_1(d(\boldsymbol{x}_1, \boldsymbol{x}_2))\qquad&\text{otherwise}.
        \end{cases}
    \end{equation*}
    Here $B_{\boldsymbol{x}_1, \boldsymbol{x}_2}$ represents the set of stream segments between the two locations $\boldsymbol{x}_1$ and $\boldsymbol{x}_2$, including the upstream segments but excluding the downstream locations, $d$ is the stream distance, $\nu$ is the nugget effect and $\omega_k$ is the weight.}

\subsubsection{Static metric graphs}
    We have already introduced graphs with Euclidean edges, termed geophysical networks, but now the time has come to define processes on them, that is: define random fields continuously indexed over edges of geophysical networks. \cite{ver2006spatial} defined proper covariance functions on (oriented) continuously indexed trees, yet the definition of covariance functions on general geophysical networks is more recent. The two main approaches used to achieve such an objective are: isometric embeddings to define isotropic covariances and stochastic partial differential equations (SPDEs).\par 
    The first approach \citep{anderes2020} exploits a classical argument \cite{schoenberg2} and embeddings in Hilbert spaces to show that the function $d\mapsto e^{-a d}$ is a valid covariance function whenever $d$ is the resistance distance. As a corollary, they obtain a whole non-parametric class of functions (termed completely monotonic) that, when composed with the resistance distance, result in valid covariance functions.\par
    Conversely, \cite{bolin_gaussian_2024} and \cite{bolin_regularity_2024} defined SPDEs whose solutions are random fields over the metric graphs. Remarkably, such an approach is able to produce random fields whose sample paths are differentiable.
    
    \begin{figure}[ht]
        \centering
        \includegraphics[width=0.7\linewidth]{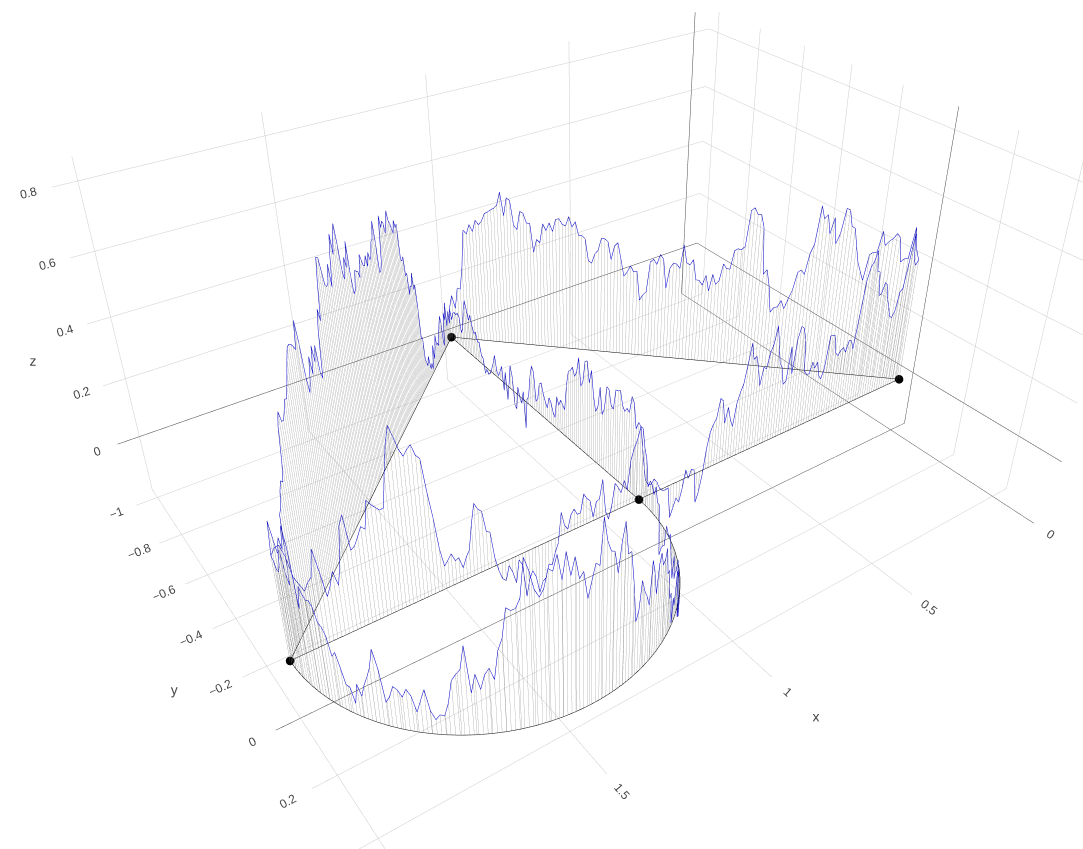}
        \caption{An isotropic process on a geophysical network.}
        \label{fig:isotropicProcess}
    \end{figure}

\subsubsection{Repeating the same metric graph over time}
     \cite{porcu2023stationary} defined a parametric class of nonseparable spatio-temporal covariance functions that allow to combine spatial and temporal measures of distance on graphs times time, for both linear and circular time. More specifically, 
    given $p$ and $k$ two positive integers, they considered two parametric classes of functions:
    \def\btheta{\boldsymbol{\theta}}
    \def\bvartheta{\boldsymbol{\vartheta}}
    \def\R{\mathbb{R}} 
    \begin{equation}
    \label{parametric_class} 
    {\cal D}_{\btheta}:= \Big \{ \varphi  (x \;  | \btheta ), \quad x \ge 0, \; \btheta \in \R^p \Big \} \,\, \text{and} \,\, {\cal H}_{\bvartheta}:= \Big \{ \psi  (t \; | \bvartheta), \quad t \ge 0, \; \bvartheta \in \R^k \Big \},
    \end{equation}
    where $\btheta$ and $\bvartheta$ are parameter vectors, and then defined
    \begin{equation} \label{eq:construction_0}
    G_{\alpha,\beta}  (x,t \;  | \btheta, \bvartheta ) = \frac{1}{\psi(t | \bvartheta)^{\alpha}}  \varphi \Bigg ( \frac{x}{\psi(t | \bvartheta)^{\beta}}\; \Big | \btheta \Bigg ), \qquad x,t \ge 0.
    \end{equation}
    Next, they provided sufficient conditions for the parametric families ${\cal D}_{\btheta}$ and ${\cal H}_{\bvartheta}$ so that
    \begin{equation*}
        G_{\alpha,\beta} \Big ( d_{\cdot}(\boldsymbol{x},\boldsymbol{x}'), |t-t'| \; \Big | \btheta,\bvartheta \Big), \qquad \boldsymbol{x},\boldsymbol{x}' \in {\cal G}\; t,t' \in \R
    \end{equation*}
    is a valid covariance function. Here $d_\cdot$ stands for either the resistance distance or the shortest path. The parameters $\alpha$ and $\beta$ have been left intentionally outside the vectors $\btheta$ and $\bvartheta$ because of their physical interpretation: when $\alpha$ and $\beta$ are both positive, $G_{\alpha,\beta}$ corresponds to a functional form that was originally proposed by \cite{gneiting2002nonseparable} for {\em space} being the $d$-dimensional Euclidean space. Several generalizations of this class are summarized in \cite{porcu201930}. When $\beta$ is positive, the spatial distance is rescaled by temporal dependence. When $\beta$ is negative, then the function acting on temporal dependence multiplies the spatial distance. \par
    The setting proposed by \cite{porcu2023stationary} allows for a substantial progress to describe processes that are dependent over both space and time. They work extremely well under the situation of a temporally {\em invariant} topology (think, for instance, of a traffic road in an old city, which is unlikely to be changed given all the constraints from building and monuments. However, there are situations where the networks are more {\em dynamical}. For instance, the traffic network in Abu Dhabi changes quite often as many parts of the city are under massive construction. For this situation, different strategies might be needed. 
\subsubsection{Covariances on time-evolving geophysical networks}
    \label{sssec:EmilioSpatioTemporal}
\begin{figure}[ht]
        \centering
        \includegraphics[width=0.9\linewidth]{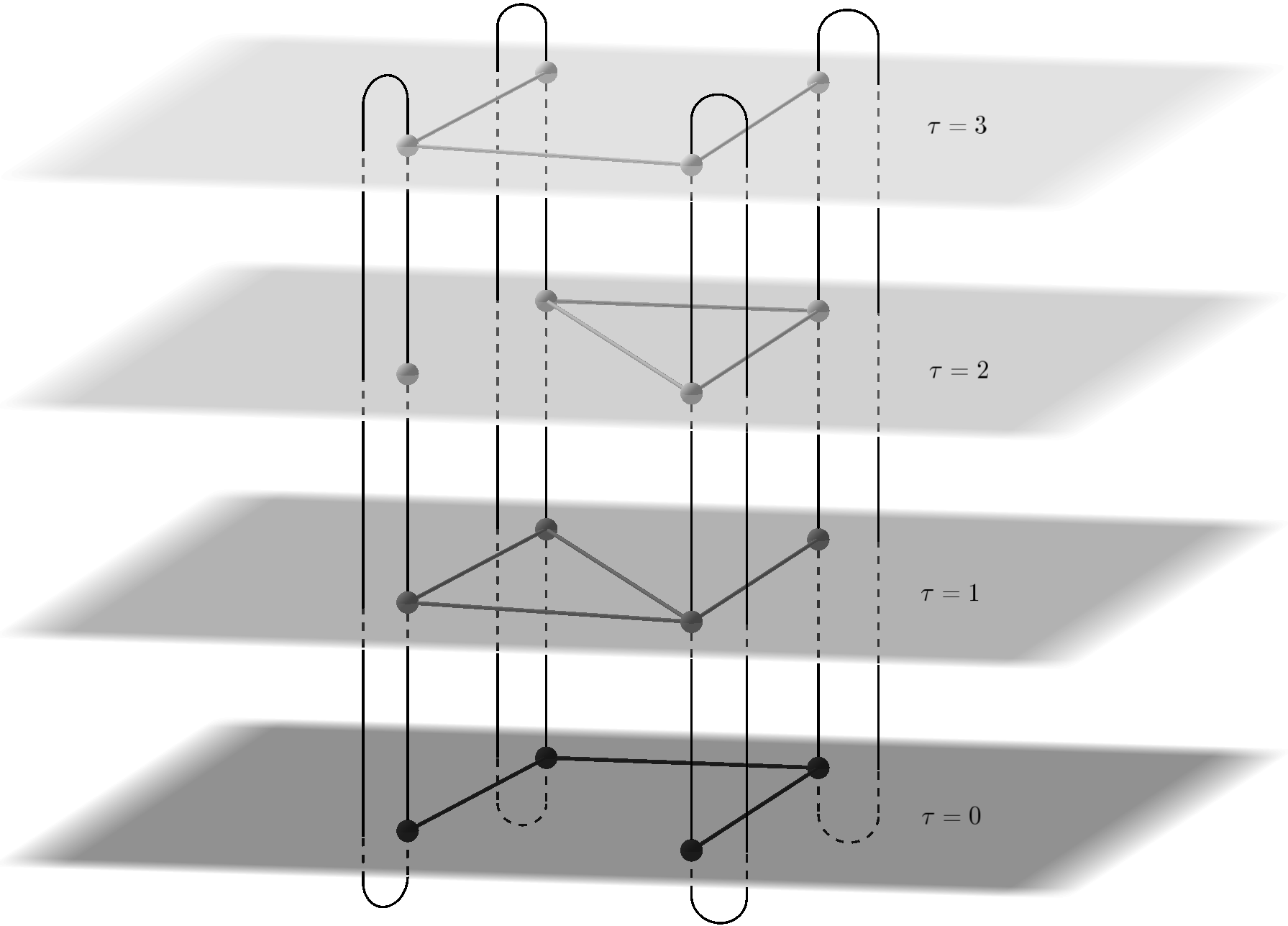}
        \caption{Construction of a time-evolving geophysical network that show a periodic evolution in time, with period $4$. The different layers ($\tau=0,\dots,3$) represent the $4$ different periodic states of the network, with ``horizontal'' edges being physical (evolving) edges, while ``vertical'' connections represent additional temporal edges that account for temporal co-dependencies. Figure adapted from \cite{filosi_temporally-evolving_2025}.} \label{fig:periodicConstruction}
    \end{figure}

\begin{figure}[ht]
        \centering
        \includegraphics[width=0.48\linewidth]{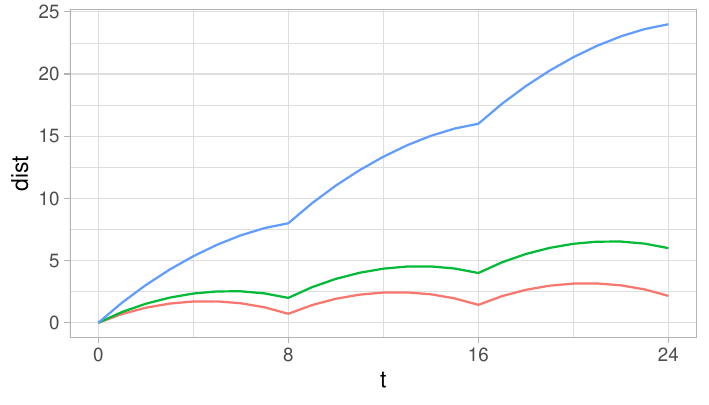}
        \includegraphics[width=0.48\linewidth]{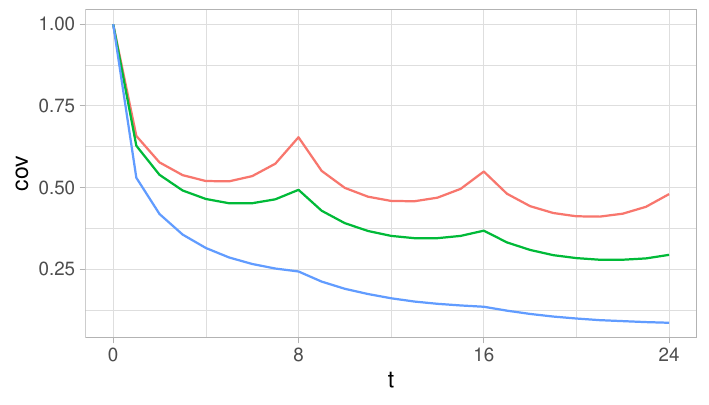}
        \caption{Example of distance functions between two vertices of (left) and the induced covariances of a process defined {\em over} (right) a periodic geophysical network of period 8 look like. The three different lines represent possible choices given by free parameters. See \cite{filosi_temporally-evolving_2025} for more details.}
        \label{fig:distancesCovariancesOnPeriodicGraphs}
    \end{figure}

    While \cite{porcu2023stationary} defined a class of covariance functions on a static graph through time, \cite{filosi_temporally-evolving_2025} formalized a mathematical framework to deal with time-evolving geophysical networks. They allow both nodes and edges to appear, disappear or change in shape and length over both linear or circular time. To cope with this complex topology, they extend the resistance distance to account for time lag and structural changes and then define a class of covariance functions that are isotropic with respect to this new distance, extending the method provided by \cite{anderes2020}. More in detail, they prove that any function 
    \begin{equation*}
        C((\boldsymbol{x},t),(\boldsymbol{x}',t')):=\psi\big(d((\boldsymbol{x},t),(\boldsymbol{x}',t'))\big),
    \end{equation*}
    where $(\boldsymbol{x},t)$ and $(\boldsymbol{x'},t')$ are spatio-temporal points on a time-evolving geophysical network, and $\psi$ belong to the special class of completely monotonic functions \citep[see, for instance,][for more details]{miller2001completely}, is a valid covariance function. Their methods are quite mathematical involved and it is out of the scope of this work to present them here. The main idea, however, is taken from \cite{anderes2020} and relies on a Hilbert space embedding of their topology. To give an intuitive insight, they consider the different shapes the graph assumes at the different time instants, thus they connect the same vertices at different times, building a new spatio-temporal geophysical network where the machinery of \cite{anderes2020} can be applied (see Figure \ref{fig:periodicConstruction} for a construction example and Figure \ref{fig:distancesCovariancesOnPeriodicGraphs} for the case of periodic graphs).    Next, they construct a proper process on the resulting spatio-temporal graph, which entails both spatial and temporal dynamics, and finally define a spatio-temporal distance as the variogram of such a process.
        
\subsubsection{Multivariate covariances on geophysical networks}
    Another extension that has been explored is the definition of multivariate processes on geophysical networks, that is: in each point of the geophysical network, \emph{many} random variables are defined. A natural example from the climate is: pressure, temperature and humidity in a given point, but other topics could have their reasons too: think about the density of different vehicles on a road network. Also this setting lends itself to the use of separable covariance functions, where one (univariate) covariance is defined over the geophysical network, while the other is defined over the set $\{1,\dots,p\}$, being $p$ the number of variables to model. \par
    \cite{filosi_vector-valued_2025} extended the Hilbert space embedding methods of \cite{anderes2020} to the multivariate case, providing a covariance (matrix-valued) function with the following properties:
    \begin{enumerate}
        \item the \emph{marginal} covariances functions coincide with the original covariances of \cite{anderes2020},
        \item the multivariate process in the vertices enjoys a sparse and intuitive conditional independence structure.
    \end{enumerate}
    In addition, they show some results useful to the construction of additional multivariate covariance functions. However, as their arguments are quite mathematical involved, we do not report them here and defer the interested reader to their paper.

\subsection{Impact on Climate}
Geophysical networks are being very popular in climate science. Here, nodes can be chosen to represent monitoring sites, hydrological junctions, sensors, or infrastructure, while edges can represent physical or functional relations, or again flows.
\cite{ver2006spatial} provided models for Gaussian processes that are defined over branching stream networks using “tail-up” and “tail-down” kernels. This work was then generalized through SSNbayes and spatial indexing frameworks in \cite{peterson2013ssnbayes, verhoef2023indexing}. Applications to river networks that are repeatedly observed over time are provided by \cite{porcu2023stationary}. \cite{sun2022gnn} developed physics-driven GNN to mimic certain hydrological processes. \cite{moshe2020hydronets} introduced the so-called {\em Hydronets} to improve runoff prediction while using the rivers peculiar topology. \par In the context of urban flood forecasting, \cite{kazadi2022floodgnn} provided {\em FloodGNN} and extended the previous work by \cite{rice2022floodgnn}. Water quality networks belong to this type of predetermined networks as well, and we mention \cite{sit2021streamflow} who proposed graph-convolutional GRUs to forecast short-term streamflow. GNN have also been used within this framework, and we refer to \cite{waterlevel2022gnn} and more recently to \cite{xie2025water_quality_gnn}. \par
Urban climate networks are, most of the time, of the geophysical type. Recently, \cite{zanfei2022graph} have used these networks to prompt GNN-based temperature prediction in urban networks. All these examples are apparently compatible with the classical geostatistical framework, see for instance \cite{verhoef2023indexing}. Thorough works that account for the river topology in concert with the choice of the right metrics can be found in \cite{verhoef2023indexing}. \par
For a comparison between traditional methods and GNN-based approaches in terms of predictive performance show non-uniform results, and the reader is referred to \cite{sun2022gnn, kazadi2022floodgnn, moshe2020hydronets}.
Transformer-based architectures (FloodGTN) use edges weighted by tide and/or dam release, showing considerable computational time gains \citep{shi2023floodgtn}.

We see applications to climate within geophysical network as four main groups: 
\begin{enumerate}
    \item Hydrologic Forecasting and River Basin Modeling: \cite{porcu2023stationary}, \cite{sun2022gnn} and \cite{moshe2020hydronets} are clear applications in this direction, in concert with \cite{zhou2022hydrograph}. 
    \item Floods and Urban Hydraulics: \cite{kazadi2022floodgnn}, \cite{rice2022floodgnn}, and \cite{rice2022floodgnn}. 
    \item Water quality: \cite{zanfei2022graph}, and \cite{suri2025powergnn}.
    \item Urban climate: \cite{li2025street_temp}, \cite{chishtie2024heatwaves}.
\end{enumerate}

\section{Networks for Climate Data}
\label{sec:networks-for}

\subsection{Why {\em for?} }
This section treats networks {\em for} climate data as tools, rather than methodological proposals as been introduced in earlier literature. We shall avoid mathematical obfuscation, while specific math can be retrieved from each of the references that we have compiled hereafter. Networks as tools can be very useful to understand who interacts with whom and at what scale. Further, tools that are typically coming from ML literature allow for qualitative assessment regarding the type of information under process. In this respect, networks {\em for} substantially differ from Tsonis and geophysical networks, as they are typically used to forecast and nowcast fields in space and time \citep{lam2023learning}, 
downscale coarse fields to local (finer resolutions) \citep{vandal2017deepsd},
  {\em denoise} irregular observations \citep{shuman2013emerging}, analyze extreme events distribution \citep{boers2019complex} and to assimilate heterogeneous observations into models \citep{evensen2009da}. \par
Some comments are in order. The example above showcases the usefulness of graphs {\em for} as pragmatic solutions to certain operational pipelines \citep{Shuman2013,Ortega2018,Kivela2014}. Moreover, the examples cited hereafter are considered as network {\em for} for three main reasons: 
\begin{itemize}
  \item The graph is instrumental to solving a task  (forecast, downscale, fill gaps), but not to claim that the edges reveal true teleconnections or a unique physical topology \citep{Shuman2013,Ortega2018}.
  \item Physics is used as a {design prior} (e.g., jet streams, drainage direction). However, edges and weights can be rather simplified or adapted for robustness and speed computation (e.g., $k$-nearest neighbor on the sphere, bipartite coarse$\to$fine) \citep{Defferrard2016}.
  \item Evaluation is {task-centric},  as the focus is on whether the graph helps prediction, reconstruction, or assimilation under regime shifts and sparsity, rather than whether its community structure or degree distribution looks the right one \citep{Rasp2020WeatherBench}.
\end{itemize}
As an opening to the descriptions following below, we stress here that there are notable differences between Tsonis networks, where edges are discovered and describe teleconnections, and networks {\em for}, where edges are chosen to route information needed to solve a given task; any resemblance to true teleconnections is incidental to the goal of better predictions or reconstructions \citep{Shuman2013,Ortega2018}. Additionally, we stress that in networks {\em for}  the topologies can be borrowed as a tool to solve a task (e.g., flood nowcasting). However, the topologies can be modified in whatever shape that can help to improve performance \citep{Kivela2014}. This is a fundamental difference with respect to geophysical networks.

As a result, the graph is the key design choice for this section. The following recipes have been widely used within climate sciences. 
\begin{enumerate}
  \item The simplest construction is that of grid or {\em mesh graph}, where one node is assigned to each grid cell and edges are chosen from the nearest geographic neighbors. Edge weights are normally assigned through inverse distance, while basic attributes such as a land–sea mask or orography provide essential context. This basic construction has nice features, such as ease of computation, and it serves as a baseline for interpolation, denoising, and forecasting tasks \citep{Shuman2013,Ortega2018,Rasp2020WeatherBench}.
\item Another popular architecture is that of
  {$k$-nearest neighbors on the sphere.} \citep{Defferrard2016,Shuman2013}. For global models (related to the whole surface of the sphere representing our planet), connecting each node to its $k$ closest neighbors by great-circle distance offers a stable (and geometry-aware) alternative to fixed stencil grids. The value of $k$ controls the local properties (neighbor structure), and the graph can be easily adapted to irregular sampling or more complicated geometries (for instance, coastlines). 
  \item In some cases, the graph structure is enriched through the addition of a small number of edges that reflect known pathways—advection corridors, such as jet streams or prevailing winds \citep{VerHoef2006,Kivela2014}.  These links can be binary or weighted, and serve as informed priors. 
\item Special graphs architectures can be used to downscaling. In this case, coarse fields and fine targets (e.g., station or high-resolution grids) are represented as two node sets,  and then each coarse node is connected to the fine nodes it influences. This technique allows to improve resolution, allows to mitigate computational burden, and keeps the mapping interpretable and modular \citep{Vandal2018,Sarkar2020}.
\item When interactions are related to variables or regions, it might be convenient to split the variables into layers, which are connected through inter-layer links \citep{Kivela2014}. This preserves within-layer geometry while enabling interaction control. 
\item The edge set might be allowed to vary over time (for instance, across seasons or ENSO phases).
For a recent treatment of this case, the reader is referred to \cite{filosi_temporally-evolving_2025} as well as an earlier contribution by \cite{Runge2019}.
\end{enumerate}

\subsection{Models}

This section discusses three families of models that have been especially influential, as networks {\em for}, within the realm of climate data science. 
\paragraph{Graph signal processing (GSP).} This family of models treats climate fields as a signal defined on the nodes of a graph: values live at grid cells or stations, and edges explain {\em who is near whom}, where {\em near} does not necessarily mean geographically. Within this setup, an important role is played by graph filters, which are linear transformations used to smooth noisy observations, and to interpolate in space-time in the presence of missing data. Regarding interpolation and denoising, applying a low-pass graph filter allows to mitigate small-scale noise while keeping coherent structures (e.g., coastal gradients or orographic bands). See for instance, \cite{Hammond2011} as well as \cite{Wang2016Trend}. 
Another important role of this model is played within the so-called {graph-based sampling}, which aims (through the same machinery) to quantify uncertainty after filtering. This amounts to evaluate nodes that might be good candidates for new sensors, which makes GSP directly actionable for network design and monitoring expansion \citep{Shuman2013,Ortega2018,Anis2016}. Finally, GSP turns to be useful for graph design purposes, rather than taking structure and topology as granted. The structure of the graph can be learned in such a way to smooth the observed signals while keeping the original data structure in terms of plausible connections \citep{Shuman2013,Ortega2018}. As a result, GSP provides a reliable baseline, and a potential prior (a building block) for more sophisticated models. \par
GSP are not Tsonis networks. To dissipate potential confusions, we note that in Tsonis networks the graph itself is the empirical result (edges represent statistical teleconnections inferred from data).
In GSP, the graph is the computational tool (given or learned as a prior) on which the climate signal is processed. Table \ref{tab:tsonis_gsp} helps for a better understanding of their differences. 
\begin{table}[H]
    \centering
    \resizebox{\textwidth}{!}{
        \begin{tabular}{|p{0.23\textwidth}|p{0.34\textwidth}|p{0.34\textwidth}|}
            \hline
            & \textbf{Tsonis networks} & \textbf{GSP frameworks} \\
            \hline
            \textbf{Graph origin} & Inferred from statistical similarity. & Predefined or learned as smoothness prior. \\
            \textbf{Role of graph} & Object of study (teleconnections). & Operator or constraint for signal processing. \\
            \textbf{Main tasks} & Diagnostics, cluster detection. & Filtering, interpolation, UQ. \\
            \textbf{Output} & Network metrics.  & Reconstructed fields and design recommendations. \\
            \hline
        \end{tabular}
    }
    \caption{Tsonis vs.\ GSP: roles of the graph.}
    \label{tab:tsonis_gsp}    
\end{table}

\paragraph{Probabilistic graphical models (PGMs).}

A complementary view of climate network connections can be achieved by looking at conditional probabilities, which allows for alternative interpretation while providing the practical computational advantage of having sparse structures, which in turn allow for scalable inference as well as for uncertainty quantification \citep{Lauritzen1996}. Within the realm of Gaussian models, sparsity regards the zeros of the precision matrix (the inverse of the covariance matrix), and the nonzeros can be interpreted as conditional links between the nodes \citep{Friedman2008,YuanLin2007,Lauritzen1996}.

PGMs provide a principled language for {spatio-temporal structure}. Dynamic Bayesian networks and related state-space formulations allow to specify lagged correlations and multi-step dependencies. Finally, PGMs represent a natural class of models for {causal discovery} \citep{lopez2022causal,  Runge2019,Shimizu2006}. No procedure yields causal truth from data alone, but PGMs at least separate the questions—\emph{is there a conditional link? which way might it point? what assumptions did we use?}—and provide uncertainty around each step. 

As previously explained, PGM can sit as an intermediate layer between GSP and GNN. The smoothness prior inherited from GSP can be used as a candidate graph architecture, which can then be refined (in terms of conditional relations) through PGM, which also provides measures of uncertainty quantification. In turn, these uncertainty measures can be embedded as {\em weights} in the loss of a GNN (on which we do not anticipate too much here). Standard references are \cite{Lauritzen1996,Friedman2008,YuanLin2007,Runge2019,Shimizu2006}.

\paragraph{Graph neural networks (GNNs).}

The third class of models learns how information should move along edges to solve a task (forecasting a field, sharpening a coarse map, or estimating a risk index for instance). GNNs are especially tailored to illustrate nonlinear relationships. Within the GNNs framework, several subfamilies have been tailored to solve specific tasks: (i) {Spectral-GNN} learns filters that can be interpreted as flexible versions of the GSP filters. For instance, Chebyshev-based convolutions and the vanilla GCN are strong baselines with few moving parts \citep{Defferrard2016}. (ii) Attention models identify dynamical connection between nodes, and are used within teleconnected regimes where the important neighbor might be far away or change with the season \citep{Velickovic2018}. (iii) Space-time models combine graph propagation with recurrence or transformers. Within this family, diffusion-convolutional RNNs and ST-GCNs are reliable choices for sequence-to-sequence forecasting, while graph transformers scale well when histories are long and exogenous drivers are many \citep{Wu2021Survey}. \par
GNNs benefit from GSP and PGMs at least for three reasons. First, a GSP filter allows for {pre-processing} and reduces noise before the GNN is applied (ensuring robustness) \citep{Shuman2013,Ortega2018}. Second, a PGM conditional network can replace a naive kNN (k-nearest neighbor) graph, focusing on edges with statistical support \citep{Friedman2008,YuanLin2007}. Finally, uncertainties from the PGM (or from an ensemble of GSP baselines) can be used as loss weights, so that the model allows for putting emphasis on regions having, for instance, sparse data, or extreme events. \par
\bigskip 
Each of these families has practical advantages, but also some caveats. 
GSP is practical and easy to implement; however, it is limited to linear relationships. PGM provides sparse and scalable structures, but heavily relies on the modeling choices (especially on the assumption of Gaussianity). Finally, GNNs represent the most flexible choice in terms of portfolio of linear relationships, but require careful regularization, monitoring for drift, and post-hoc explanations; they are also {\em hungry} for consistent training data \citep{Wu2021Survey}. Used together, the three families cover the spectrum from simple and stable to rich and adaptive while preserving a clean path for uncertainty and governance.
Table \ref{tab-class} illustrates the use of this methods, according to the assigned task, within the realm of climate models. 

\begin{table}[t]
    \small
    \setlength{\tabcolsep}{4pt}
    \centering
    \resizebox{\textwidth}{!}{
        \begin{tabular}{|>{\raggedright\arraybackslash}p{3cm}|p{5.8cm}|p{6.8cm}|}
        \hline
        \textbf{Task family} & \textbf{Graph recipe} & \textbf{Models} \\
        \hline
        {Global and regional forecasting} &
        kNN on great-circle distance; add a few physics-aware links; multilayer across variables or levels &
        Spatio-temporal GNNs (GCN/attention) with temporal GRU/transformer heads; robust to missing data. \\

        {Downscaling and super-resolution} &
        Bipartite coarse$\to$fine graph with distance, orography and land-cover attributes &
        Graph U-Nets and cross-scale attention; preserve sharp fronts and orographic effects \citep{Vandal2018,Sarkar2020}. \\

        {Imputation and gap-filling} &
        Sensor or mesh graph; edges from rolling similarity and distance &
        Combine a simple graphical model prior with a GNN denoiser; train with masked losses \citep{Friedman2008}. \\

        {Extremes and compound events} &
        Event graph built from threshold exceedances; edges capture lag or tail dependence &
        Detect backbones and hubs during heatwaves and atmospheric rivers \citep{Boers2019}. \\

        {Hydrology and floods} &
        Directed river network; weights informed by slope and travel time &
        Physics-informed GNNs; Graph Transformers with conservation penalties \citep{VerHoef2006}. \\

        {Urban microclimate} &
        Heterogeneous graph over roads, buildings, and sensors &
        Heterogeneous GAT; mitigation scenario testing \citep{Chishtie2024}. \\

        {Data assimilation} &
        Grid graph with observation--to--state links &
        Use graph-aware covariances inside EnKF/4D-Var; keep physics in the analysis step \citep{BocquetSakov2013}. \\
        \hline
        \end{tabular} 
    }
    \caption{\label{tab-class} Graph design patterns for common networks {\em for} tasks.}
\end{table}

The constraints that are proper of each of these three families should not alter the physics properties. A couple of principles are listed below. \par
\begin{enumerate}
    \item When predicting fields over a geophysical network, constraints should preserve certain structures. On rivers, {\em inflow minus outflow equals storage change}, and a similar principle applies to atmospheric budgets \citep{BocquetSakov2013}. \par
    \item The physics can be related not only to the geophysical network, but also to the data structure. Some data must be non-negative (precipitation, runoff) or live in sensible ranges \citep[relative humidity, wind speed, see][]{BocquetSakov2013}. Another relevant situation is related to the presence of coastlines, boundaries and barriers, and the use of weights should reflect smoothness in the interior while allowing for sharp transitions in persistent fronts. This principle has been analyzed for both graph modeling and downscaling, and the reader is referred to \cite{Wang2016Trend,Shuman2013,Ortega2018} for the former, and to \cite{Vandal2018} for the latter.
\end{enumerate}
\bigskip \par
Uncertainty quantification plays an important role in networks {\em for}. Within a huge literature, we mention here (without being exhaustive) deep ensembles \citep{Lakshminarayanan2017} and scenario-aware checks, for which it is verified when and where uncertainty widens in unfamiliar regimes (e.g., unusual ENSO phase).

\section{Bridges {\em within} the Triad}

\subsection{Discrete {\em versus} Continuum Hypotheses}

An intuitive objection against the approaches presented in this paper is represented by their potential redundancies in front of the well-established continuous space--time statistical models \citep{porcu2023mat, porcu201930}. The review by \cite{porcu201930} provides a thorough account of continuous space-time models, with special emphasis on the Gaussian case, starting from models on the sphere into spectral models and models based on SPDEs \citep{lindgren2011explicit}. This rich literature has provided a wealth of contributions to climate modeling in terms of interpolation, prediction, uncertainty quantification, and physically-informed inference across the planet’s surface \citep{cas17}. An understandable criticism is oriented towards the fact that Tsonis networks 
 lack theoretical foundations, and introduce discretization techniques that might be easily avoided through the use of continuous models. We do not necessarily agree with this statement. In particular, we claim the following. \par
 \paragraph{Approaches are complementary.} While space-time models are generative, Tsonis networks are inductive. Hence, they do not assume that the data come from a {pre-defined} stochastic process (for instance, a Gaussian process). From a epistemological viewpoint, both approaches are needed: the first explains, the second detects. 
 \paragraph{Abrupt Changes.} Continuous models are at present struggling to detect abrupt changes, that are structural transitions and nonstationarities. Typical smoothness assumptions can actually mask rapid changes and restructurings in climate dynamics. On the other hand, networks are excellent in detecting transitions.  These topological shifts are interpretable as early warning signals and are less visible in classical geostatistical outputs.
\paragraph{Resilience.} Space-time models require a sufficiently dense data coverage and are normally based on strong assumptions ({\em e.g.}, stationarity). However, for most of the planet the available observations are very sparse. Climate networks are more resilient to such limitations, as they include statistically significant links only. Further, they enable inference even in irregularly-sampled domains.
\paragraph{Policy Oriented.} This aspect will be expanded in the subsequent section. Governance decisions often require communication between technical experts, decision-makers, and the public. Narratives centered on “central nodes”, “broken links”, “fragmentation”, “resilience” are naturally accepted by policy and decision makers.

We do not advocate abandoning continuous models, but rather integrating them with network approaches. Graphs can inform the design of spatial priors, guide the identification of anisotropies, and contribute to kernel shaping or nonstationary extensions. Likewise, continuous models can be used to simulate or stabilize network estimation under uncertainty. Hybrid frameworks, such as multi-layer graphs or coupled GNN-GP architectures, promise rich avenues for uniting statistical power with topological insight.

\subsection{Different Worlds}
We start with an epistemic discussion that concerns the possible types of networks opened by the bridging {\em within} philosophy proposed herein. We consider here a combination of some Tsonis networks as well as scale networks, with the geophysical networks. Table \ref{tab:topology_framework} gives a reliable picture.
\begin{table}[ht]
    \centering
    \small
    \resizebox{\textwidth}{!}{
        \begin{tabular}{|p{3cm}|p{5.0cm}|p{6cm}|}
            \hline
            \textbf{Network Type} & \textbf{Inferred from Data} & \textbf{Given by Domain Knowledge} \\
            \hline
            Random-like 
            & Correlation networks at low thresholds \citep{tsonis2006networks}.
            & Sensor networks or monitoring grids without structure. \\
            Small world 
            & Mutual information networks \citep{donges2009complex}. 
            & River networks with man-made channels \citep{muhammad2016managing}. \\
            Scale free 
            & ENSO-related hubs in dynamic networks \citep{tsonis2008topology}.
            & Electric or road networks in urban heat island studies \citep{schafer2018dynamically}. \\
            \hline
        \end{tabular}
    }
    \caption{Framework: Topology $\times$ Construction Mode.}
    \label{tab:topology_framework}
\end{table}

We provide a description of the {\em interesting} ones as those networks stemming from the following combinations:
\begin{enumerate}
    \item Random Networks can have a predetermined structure. For instance, sensor networks for which the nodes have been established without invoking any precise organization principles. Typical situations are those of climate or environmental monitoring stations \citep{mearns2003climate}. 
    \item Small world networks can also have a predetermined topological structure. Hydrological networks \citep{ver2006spatial}, in particular those having engineered channels, can exhibit the characteristics of a small world network.

    Some geophysical networks can be scale-free. For instance, urban infrastructure or road systems often follow scale-free distributions due to cost-efficiency and resilience trade-offs. Their climate relevance lies in vulnerability analysis under extreme weather events \citep{schafer2018dynamically}.
\end{enumerate}

This framework allows researchers to integrate domain-specific knowledge with statistical inference, facilitating interdisciplinary approaches and hybrid models that respect both empirical data and physical constraints.


\subsection{Technical Bridges}
    This section provides some ideas for a transition from separate paradigms into potential fusions. Providing bridges is not a mere intellectual exercise, but opens for a much wider discussion that has been in the core of data science in the last years: should data science be based on {\em data-first} or {\em model-first} approaches?

\subsubsection*{Graph-Signal Reweighting of Tsonis Networks}

We have already discussed the intrinsic drawbacks in providing binary large scale connections through Tsonis networks. The recent work by \cite{haas2023pitfalls} provides an excellent insight into the statistical limitations of such an approach. It might be tempting to integrate the philosophy of signal processing inside the Tsonis architecture --- hence, graph signal processing \citep{Shuman2013, Perraudin2017} might represent an interesting tool for integration. Let $G = (V, E, W)$ denote the Tsonis network, where $V$ represents nodes (spatial locations), $E$ the edges (pairs with indicators of statistical similarity being above a given threshold), and where $W$ the adjacency weight matrix. We consider the {\em climate signal} as the process defined over the nodes at different time instants. The idea is (a) implement a Tsonis network, (b) consider the process defined over the Tsonis network as a graph signal, and (c) update the weighting matrix $W$ with a matrix $\widetilde{W}$ based on graph signal similarity. (As an alternative, one might use graph-based smoothing through a graph Laplacian regularization approach.) The optimized weights for $\widetilde{W}$ can be iteratively adapted to minimize temporal inconsistency, providing a data-driven refinement of the edge structure \citep{Thanou2017, Segarra2016}. This operation can be evaluated {\em locally} by embedding, {\em e.g.}, a urban network into the pre-existing Tsonis network. Then, the restriction to a subgraph corresponding to the Tsonis graph at the region where the urban network is evaluated, one can create hybrid architectures.

Finally, the reweighted network can be analyzed via the graph Fourier transform to identify coherent modes of variability \citep{Shuman2013, mendes2014signal}. This may reveal latent mesoscale phenomena not observable in the original Tsonis formulation. This integration of dynamic graph signal characteristics with climatological correlation networks provides a promising path to unify the ``of'' and ``over'' paradigms.

\subsubsection*{Local Embedding of Metric Graphs into Climate Teleconnections}
Spatially-constrained metric graphs can be embedded into abstract teleconnection networks. We shall avoid mathematical obfuscation and will instead focus on the ideas for such embedding strategies. 
We start with a global graph $G_{\text{Glob}}$ of the Tsonis type, and with a local metric graph $G_{\text{Loc}}$ equipped with some metric $d$. To embedd $G_{\text{Loc}}$ into $G_{\text{Glob}}$, we follow the steps. (a) Assign geographical coordinates to the nodes of $G_{\text{Loc}}$. This can be done through remote sensing, for instance. Then, each node is assigned to a single node or a cell, within $G_{\text{Glob}}$. (b) Downscale the climate signals available at $G_{\text{Glob}}$ into $G_{\text{Loc}}$, for instance through kriging or through graph based filtering 
\citep{Perraudin2017, Dong2019}. (c) Decide on modeling strategies, where the two graphs might be treated as nested, or as multilayers. Cross-scale interactions can be modeled using the available techniques such as graph alignment or multi-resolution techniques \citep{Kivela2014, DeDomenico2013}. A typical examples is that of urban temperatures (heat mitigation), where the micro-drivers, such as ENSO or Indian monsoon, are located over a global Tsonis network. These approaches are very promising to perform local inference, as well as to propagate information between two different scales \citep{Kipf2017, Bronstein2021}. 

\subsubsection*{Climate-Informed Graph Topology Learning}
    Previous approaches where based on mutual embedding. Here, we suggest some procedures to {\em learn} the graph topologies starting from climate data. This idea has been partially pursued by \cite{Dong2019} within the realm of graph topology learning. Adapted to our context, this is about inferring topology of networks in such a way that these reflect the dynamical dependencies while respecting the geographical constraints. This class of approaches is typically based on finding a graph Laplacian, or and adjacency matrix, such that the signal is smooth over the graph, constrained on some functional minimization problem. One might object that such approaches should be strictly connected to the physics nature of climate data. This might be taken into account by, {\em e.g.}, penalizing connections that contradict known geophysical distances (this can be implicit in the type of kernel that has been used). Bayesian approaches might do a good job through the design of priors that incorporate causality structures. Finally, multivariate based-process techniques might add information on the existence of edges.  As a practical example, consider performing inference of a regional graph over the Middle East, using precipitation and temperature data from satellite and reanalysis products. \par
    These bridges motivate a unifying formalism involving multi-layer networks \cite{Boccaletti2014} or more sophisticated form of graph associations \citep{DeDomenico2013,Berlingerio2011}. 

\subsection{Epistemic and Governance-based Bridging {\em within}} 
We have analyzed the possibility of having hybrid systems that integrate dynamical coherence with georeferenced constraints of the climate system. Such an integration can be viewed from the angle of climate governance, as much as we have done for the case of Tsonis networks. \par
When Tsonis-style empirical graphs are aligned with predetermined network geometries, they produce hybrid structures that embed observed dynamical dependencies within structural or jurisdictional boundaries. Hence, such an integration will improve capabilities in terms of prediction accuracy, opening the door to multi-level decision making. \par
In view of networks integration, early warning systems can be improved by enabling the detection of teleconnections under the presence of physical constraints. For example, a Tsonis-derived correlation between upstream precipitation in Region A and downstream flood occurrence in Region B can be embedded onto a river basin network and improve risk predictability. This would be coherent with the purported anticipatory protocols suggested by both national and international agencies. \par
Hybrid networks can become successful to integrate information within different administrative boundaries ({\em e.g.}, the provinces in Spain). This fact can activate trans-boundary governance agreements.\par
Within the realm of urban planning, the identification of climate-derived teleconnections and their embedding on urban networks may allow municipalities to identify the so-called {\em compound hazards} (for instance, simultaneous blackouts) that may happen within correlated neighborhoods. \par
We elaborate a bit more within these directions, wih emphasis on increasingly polycentric and data-driven decision-making frameworks. \par
\paragraph{From Diagnostic to Design-Oriented Governance.} Network models that integrate Tsonis-style empirical structures with jurisdictional and geophysical constraints align naturally with emerging paradigms of anticipatory and risk-informed governance. The integration allows to detect both teleconnections and cascading risks and in turn prompt proactive attitude to enable scenario planning that respects both physical and institutional boundaries \citep{ostrom2017polycentric}. \par
\paragraph{Climate Finance, Risk Pools, and Sovereign Instruments.} Fusion-based networks can inform climate-indexed financial instruments, including catastrophe bonds and sovereign risk pools. For example, parametric insurance models have been proposed on the basis of metrics (based on networks) of teleconnected droughts or flood propagation patterns \citep{hellmuth2009index}. The African Risk Capacity \citep{clarke2013cost} and the Caribbean Catastrophe Risk Insurance Facility \citep{ghesquiere2006caribbean} have both underscored the need for hazard models that reflect structural linkages and localized vulnerability. \par
\paragraph{Calibration within International Frameworks.} International connections between different policies and paradigms can be favored by networks integration. The Sendai Framework for Disaster Risk Reduction is centered on improving risk governance through multi-hazard early warning systems and disaggregated risk metrics \citep{united2015sendai}. Hybrid networks can favor the embedding of risk metrics within physical and administrative systems. Analogously, the Task Force on Climate-related Financial Disclosures strongly favors explicit metrics of spatial exposure and systemic risk, which can be enriched by network-based reasoning \citep{tcfd2017final}. \par
\paragraph{Polycentric Adaptation.} Climate adaptation increasingly occurs within decentralized, overlapping jurisdictions. Multilevel coordination and improvement of resource allocation, as well as accountability, can be all improved by embedding empirical dependencies into networks that align with local governance geometries \citep{jordan2018governing, ostrom2017polycentric}.  \par
We do recommend a development of governance in accordance with the integration of the two networks, especially through multi-layer governance approaches, and through the use of methods that will allow to redefine exposure and vulnerability metrics across multiple governance units.


\section{The Shaw--Stevens Dialogue and the Fusion {\em Between}}

This section proposes a dialogue between the Shaw-Stevens agenda and the network triad. We start by recognizing the different genealogies of the two paradigms - the former being largely inspired by the philosophy of climate science, and the latter substantially coming from data-centric and computational frameworks. \par
However, they share dissatisfaction for LSD, and this fact is made explicit within the Shaw-Stevens discourse while it reads between the lines of the triad proposal in this paper. This section proceeds by outlining the contrasts between the two paradigms, to then open for a reconciliation that will allow for a transition to what we term, throughout, a fusion {\em between}.

\subsection{Contrasts and Synchronies {\em Between}}

According to the LSD framework, the Earth system is represented on a latitude–longitude grid.  Sub-grid processes are parameterized in terms of grid variables, and local properties are defined with respect to the Euclidean metric.  Hence, neighbor structures are determined by those nodes that lie next to one another on this discretized (ad flat) version of the globe.  In the network triad, adjacency is not necessarily given or determined through a simplified version of geographical proximity. For Tsonis networks, proximity is relational. Hence, there is a clear ontological shift - from a uniform, rectilinear grid to a flexible graph whose geometry adapts to the processes under study \citep{Donges2009,Boers2019}.  Distance ceases to be geographical and becomes contextual for Tsonis networks. It is geographical for geophysical network, but the metric is carefully chosen according to the technical developments illustrated in previous sections. \par
There is an apparent epistemic divergence as well. LSD is generative, while Tsonis networks are inductive. Within geophysical networks, there is certainly space for generative approaches, while networks {\em for} can be of any type (generative, predictive, inductive). \par
Another potential discrepancy stands with the fact that \cite{ShawStevens2023} criticize ML as disruptive but opaque.  Our approach embraces ML but mitigates it with interpretable layers. These distinctions are summarized in Table~\ref{tab:interpost}. 
\begin{table}[H]
    \centering
    \resizebox{\textwidth}{!}{
        \begin{tabular}{|p{0.28\textwidth}|>{\raggedright\arraybackslash}p{0.32\textwidth}|>{\raggedright\arraybackslash}p{0.32\textwidth}|}
        \hline
        \textbf{Dimension} & \textbf{Standard Approach (LSD)} & \textbf{Network Triad (of/over/for)} \\
        \hline
        Ontology of space & Gridded, Locally Euclidean  & Relational, geodesic, anisotropic \\
        Epistemology & Generative, dynamical closure & Inductive, detective, relational diagnostics \\
        Role of ML & Disruptive but opaque & Physics-informed, interpretable via graphs \\
        Evidence mode & Hypotheses via canonical models & Hypotheses via network statistics and edge tests \\
        \hline
        \end{tabular}
    }
    \caption{Contrasting the ``Standard Approach'' of Shaw--Stevens (LSD) with the Network Triad.}
    \label{tab:interpost}
\end{table}
Beneath these contrasts lie profound synchronies.  \citeauthor{ShawStevens2023} call for hierarchies --- that is, from simple to complex ladders of models, mechanism-denial experiments, and strategic nudging of dynamics.  The network triad provides a natural tool for such hierarchies.  In fact, Tsonis networks can be used as a baseline, the geophysical as an intermediate layer (through domain knowledge embedding), and 
networks \emph{for} climate data represent the upper layer, where algorithms exploit graph structure to perform forecasting, assimilation, or downscaling.  \par
There is an apparent convergence between the two paradigm in the way discrepancies are embraced. \cite{ShawStevens2023} urge climate scientists to embrace mismatches  (Southern Ocean cooling paradox,  Walker-circulation slowdown, or terrestrial humidity declines) as an indispensable source of data discovery rather than {\em noise} to be eliminated from the trend. Within the network triad, discrepancies come as measurable shifts in graph statistics: changes in community structure,
degree distribution, or causal orientation. \par
As a final remark, \citeauthor{ShawStevens2023} call for disruptive computation (very large ensembles, kilometer-scale models, ML) which is in apparent synergy with our network agenda.  Each direction proposed in this paper implies a computational disruption, but we also insist on keeping interpretable structures.
In this sense, both programs seek acceleration without opacity.

\subsection{The Fusion {\em Between}}
Our narrative is pointing to the fusion {\em between} as a natural step forward. While Shaw-Stevens provides the vision and mission of climate science for the years to come, the networks triad provides the concrete artillery to realize that vision and mission.  The natural fusion discussed here is illustrated below through some examples. \par 
Hierarchies are indispensable for identifying the limits of LSD, and the triad provides very good tools for such a task. In particular, through this lens, LSD becomes a hypothesis that can be explicitly tested within a relational framework. \par 
\citeauthor{ShawStevens2023} recommend that discrepancies should drive observational design rather than being smoothed away.  Network formulations make this actionable:
uncertainty fields can be defined on nodes and edges, informing adaptive sampling strategies that target regions of maximal epistemic tension. Table \ref{table:between} illustrates these connections through practical examples.

\begin{table}[h] 
    \centering 
    \resizebox{\textwidth}{!}{
        \begin{tabular}{|p{0.28\textwidth}|>{\raggedright\arraybackslash}p{0.32\textwidth}|p{0.32\textwidth}|}
            \hline
            \textbf{Discrepancy} & \textbf{Standard Approach (LSD view)} & \textbf{Network Remedy} \\
            \hline
            Walker-circulation slowdown & Attributed to large-scale ocean–atmosphere coupling; parameter uncertainty & Causal graphs orient edges; PCMCI and LiNGAM identify mediators and exclude passengers \\
            Land-humidity decline & Expected closure with temperature; models diverge regionally & Graphs over basins capture soil–atmosphere coupling; sensor placement reduces uncertainty \\
            Southern Ocean cooling & Models struggle with mixed-layer depth and fluxes & Graph geodesics along isopycnals refine assimilation; targeted observations via community detection \\
            Storm-track biases & Parameterized small-scale eddies in GCMs & Cross-scale bipartite graphs conserve fluxes while preserving coastal/orographic bands \\
            \hline
        \end{tabular}
    }
    \caption{Examples of regional discrepancies and corresponding network remedies.}
    \label{table:between}
\end{table}
The fusion extends beyond science into governance.  Networks provide the metrics that can be used to compare different ensembles and then to implement validation, communication and decision making. \par
In conclusion, the two agendas are not antagonistic but complementary.

\section{The Meta Fusion: the Shaw-Steven Network Ecosystem}
The intellectual trajectory of this paper culminates in what may be called a \emph{fusion within the fusion}.  We have explored the inner fusion, termed {\em within}, and built a unique framework for empirical connectivity, algorithmic complexity, and spatial geometry $\&$ topology. We have also bridged this triad with the epistemological agenda proposed by \cite{ShawStevens2025}, whose essay on the ``Other Climate Crisis'' questions the dominating paradigm of LSD. There is way more than the illustrated parallelism between the two frameworks, and finding suitable connections requires resorting to philosophical arguments as a baseline for scientific thinking. We argue below that the network triad is not just the tool for the implementation of the Shaw-Stevens agenda, but a {\em sine qua non} element for what we term a Shaw-Stevens network ecosystem, that is, this section argues that a {meta fusion} between the two frameworks creates a system where epistemology and methodology co-produce one another. Our argument is supported by the following reasonings. 

\subsection{Reflexive Closure}
We start by arguing that the network triad and the Shaw-Stevens agenda, while different in their ontological and epistemic baselines, converge through their interaction. Specifically, we argue that their interaction reveals a deeper principle: {\em the architecture of knowledge in climate science is itself a network}. By this principle, we mean that the architecture of knowledge in climate science behaves under the same logic of self-organization and feedback that governs the climate system itself, where the data universe, information, models and discoveries become a network having feedback loops that allow for the network to dynamically re-organize itself. Let us carefully justify our assertion. \par 
A first alignment comes from the fact that at the methodological level, the triad provides a grammar of relations. Tsonis networks capture empirical coherence; networks {\em for} translate this coherence in tools for inference, and geophysical networks position the previous one in a geometric continuum, connecting statistics and computation to the topologies of our planet. This first alignment provides duality for each component of the triad, where each of them is boosted by the others while being at the same time a condition for their coherence. \par
A second remark regards alignments within the Shaw-Stevens agenda. Parameterizing the {\em small through the large} is not necessarily  a good idea anymore. It is instead time for a reflexive science, where learning comes from interaction of scales rather than from their separation. The network triad is a natural grammatical integration within this narrative, as the triad network is the perfect candidate for multiscale reasoning. Hence, our second alignment illustrates how the network triad becomes a tool for verifying the Shaw--Stevens hypothesis in practice. \par
We culminate this by invoking what we term a {\em reflexive closure}, which is substantially the meta fusion emerging from the interaction of the two previous alignments. Hence, the network triad becomes a philosophical baseline that goes beyond providing tools for the Shaw-Stevens program. At this stage, the construction of networks becomes self-validating in the sense that it provides its own epistemic tests by revealing when the statistical relations among data confirm or contradict the modeling assumptions that motivated them. Conversely, the Shaw-Stevens framework, once an abstract critique of scientific culture, becomes operational. In particular, it informs choices of threshold, regularization, and hierarchical design in network modeling. We clarify that reflexivity is achieved when a system has categories, classes and variables that are inseparable but keep their distinct identities.

\subsection{Morphology of the Meta Fusion}
By morphology we mean the architecture stemming from the integration of methods and epistemology within the meta fusion. While reflexive closure invokes hierarchies to then build a circular system, this section insists on the circulation part to argue in favor of an ecosystem that dynamically reorganizes itself. \par 
A simple way to look into this is through the following layers: 
\begin{enumerate}
    \item Internal coherence within the triad, which is the basis for the invoked fusion {\em within}. Internal coherence is the condition for internal closure as previously advocated;
    \item External alignment with epistemology, which allows the triad to transcend instrumentalism;
    \item Reflexive closure to determine autopoiesis, that is the system becomes capable of conditions of its own validity.
\end{enumerate}
A simplified version of the morphology above is provided through Table \ref{tab:metafusion}.

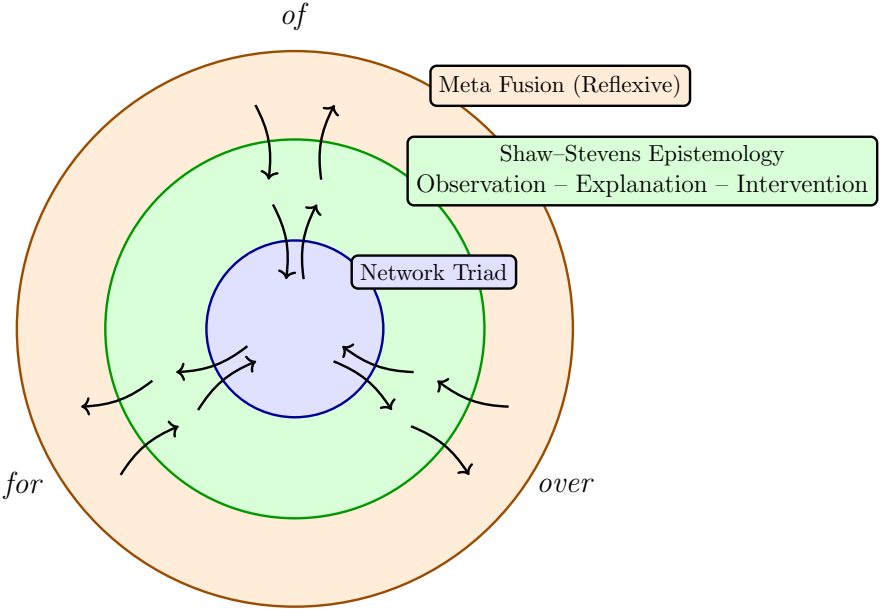
\begin{figure}[H]
    \centering
    \scalebox{0.8}{
        \begin{tikzpicture}[scale=1.05, every node/.style={align=center}]
            \def\rinner{1.4}
            \def\rmid{3.0}
            \def\router{4.4}
            \def\arrowDeltaRadi{0.6}
            
            \tikzset{
                layerlabel/.style = {draw, rounded corners=3pt, thick, fill=white, inner sep=4pt, line width=0.4mm},
                arcarrow/.style   = {->, thick, black, line width=0.4mm},
            }
            
            \fill[orange!15, draw=orange!60!black, thick, line width=0.4mm] (0,0) circle (\router);
            \begin{scope}
                \clip (0,0) circle (\rmid);
                \fill[green!15] (-\router-1,-\router-1) rectangle (\router+1,\router+1);
            \end{scope}
            \draw[green!60!black, line width=0.4mm] (0,0) circle (\rmid);
            \fill[blue!12] (0,0) circle (\rinner);
            \draw[blue!60!black, line width=0.4mm] (0,0) circle (\rinner);
            
            \node[layerlabel,fill=orange!15] at (4.2, \router-0.55) {\small Meta Fusion (Reflexive)};
            \node[layerlabel,fill=green!15] at (5.5,  \rmid-0.5)     {\small Shaw--Stevens Epistemology\\ Observation -- Explanation -- Intervention};
            \node[layerlabel,fill=blue!12] at (2.2, \rinner-0.5)           {\small  Network Triad};
            
            \node at (90:\router+0.55)  {\large \emph{of}};
            \node at (210:\router+0.55) {\large \emph{for}};
            \node at (330:\router+0.55) {\large \emph{over}};
            
            \foreach \ang in {80,200,320}{
                \draw[arcarrow, black]
                (\ang:\rinner-\arrowDeltaRadi) to[bend left=18] (\ang:\rinner+\arrowDeltaRadi);
                \draw[arcarrow, black]
                (\ang:\rmid-\arrowDeltaRadi)   to[bend left=18] (\ang:\rmid+\arrowDeltaRadi);
                \draw[arcarrow, black]
                (\ang+20:\rmid+\arrowDeltaRadi) to[bend left=18] (\ang+20:\rmid-\arrowDeltaRadi);
                \draw[arcarrow, black]
                (\ang+20:\rinner+\arrowDeltaRadi)   to[bend left=18] (\ang+20:\rinner-\arrowDeltaRadi);
            }
            
            
        \end{tikzpicture}
    }
    \caption{Meta-fusion as three nested layers. The inner \emph{triad} (of--for--over) sits within the Shaw--Stevens epistemology, itself enclosed by a reflexive meta-layer. Red arrows indicate two-way feedback: methodology shapes epistemology and vice versa.}
    \label{fig:metafusion}
\end{figure}

\begin{table}[H]
    \centering
    \renewcommand{\arraystretch}{1.8}
    \scalebox{0.82}{
    \begin{tabular}{|p{0.22\textwidth}| p{0.26\textwidth}|p{0.26\textwidth} |p{0.26\textwidth}|}
    \hline
    \textbf{Dimension} &
    \textbf{Fusion within the Triad} &
    \textbf{Fusion with Shaw--Stevens} &
    \textbf{Meta-Fusion (Reflexive)} \\
    \hline
    \textbf{Ontological} &
    Data, model, and domain are unified into a single representation.  The triad triggers an ontology of relations where there is no more distinction between object and context. &
    From processes to paradigms: the triad supplies a concrete ontology compatible with the Shaw-Stevens agenda. &
    Mutual embedding of data ontology and epistemology.  The network becomes a co-agent in data universe updates. \\[0.6em]
    \textbf{Methodological} &
    Inference, geometry, and observation co-evolve.  Empirical and generative logic fold into one another through network dynamics. &
    Hierarchical constructions test and inform the critique of LSD. Network hierarchy offers a tool for plural and adaptive modeling. &
    Recursive design of hypotheses: methodology interrogates its own assumptions.  Epistemology becomes a design principle guiding computational practice. \\[0.6em]
    \textbf{Ethical / Operational} &
    Transparency and interpretability are the essence of model design; interpretability and trustworthiness acts as moral constraints within computation. &
    Accountability and responsibility emerge as conditions for explanation. Hence, ethics becomes an epistemic dimension. &
    Co-governance of models and meanings.  Data ethics and epistemic humility are folded within methodological loops, ensuring that closure remains open to critique. \\
    \hline
    \end{tabular}}
    \caption{Schematic synthesis of the three levels of fusion: within the triad, between the triad and the Shaw--Stevens paradigm, and their reflexive meta-fusion. Each row corresponds to an ontological, methodological, or ethical dimension of interaction.}
    \label{tab:metafusion}
\end{table}
\noindent Horizontal reading of the table illustrates the integration of knowledge layers: from technical synergy, to epistemic alignment, to reflexive autonomy.  The vertical axis instead describes the intensity of engagement between the three levels. 
Taken together, the three columns and three rows define a simple scheme that describes the morphology of the meta fusion we have in mind. Figure \ref{fig:metafusion} provides a geometric simplification for this closure.

\subsection{Autopoiesis of Climate Knowledge}
The culmination of the invoked fusion is an image of climate science that is autopoietic---that is, a system capable of producing and renewing its own conditions of possibility. We borrow this term from the exemplary theory from Maturana and Varela’s \citep{MaturanaVarela1980} of living systems. For the authors, an autopoietic system is capable of maintaining itself through the continuous reproduction of its components. \par
Our claim of autopoiesis of this fusion is not exceptional by itself, as it is consistent with climate system, being also autopoietic. In facts, climate systems reproduce its statistical regularities through the continual circulation of matter and energy. Further, climate systems achieve stability not by eliminating disequilibrium but by continually converting gradients and perturbations into patterns of renewal-circulations that restore balance through motion.  \par
Accordingly, our approach claims for a construction of an autopoietic climate science that transforms uncertainty into insight, and insight back into uncertainty. \par
We also claim that the network triad becomes a data agent within the autopoietic system, with a task of generating knowledge through data discoveries that update the state of the data universe. \par
We stress that autopoiesis should not be taken as solipsism. For \citeauthor{MaturanaVarela1980}, a living system maintains its identity by remaining open to energy and information from the environment. We invoke the need to the same principle to apply inside the proposed ecosystem that finds its essence in the engagement with social,
political, and ethical realities. \par
In this meta fusion, closure and openness are no longer opposites but phases of a single process.

\bibliographystyle{apalike}
\bibliography{bib}

\end{document}